\documentclass[twocolumn,tighten,times,floatfix]{aastex631}
\usepackage{amsmath}
\usepackage{anyfontsize}
\usepackage{amssymb}
\usepackage{sidecap}
\usepackage{natbib}
\usepackage{lipsum}
\usepackage{nameref}
\usepackage[T1]{fontenc}
\usepackage{graphicx}
\usepackage{hyperref}
\usepackage{svg}
\usepackage[normalem]{ulem}
\usepackage{soul}

\newcommand{\Kepler}{\textit{Kepler}}
\newcommand{\tess}{\textit{TESS}}
\newcommand{\kepler}{\textit{Kepler}}

\newcommand\blfootnote[1]{%
  \begingroup
  \renewcommand\thefootnote{}\footnote{#1}%
  \addtocounter{footnote}{-1}%
  \endgroup
}

\shorttitle{SN2023bee}
\shortauthors{Wang et al.}
\graphicspath{{./}{figures/}}

\begin{document}

\setlength{\abovedisplayskip}{3pt}
\setlength{\belowdisplayskip}{3pt}

\DeclareRobustCommand{\ori}[1]{ {\begingroup\textcolor{orange}{Ori: #1}\endgroup}}

\newcommand{\JHU}{Physics and Astronomy Department, Johns Hopkins University, Baltimore, MD 21218, USA}
\newcommand{\JHUCS}{Department of Computer Science, Johns Hopkins University, Baltimore, MD 21218, USA}
\newcommand{\STScI}{Space Telescope Science Institute, Baltimore, MD 21218, USA}
\newcommand{\NCPAS}{National Centre for the Public Awareness of Science, Australian National University, Canberra, ACT 2611, Australia}
\newcommand{\UCSC}{Department of Astronomy and Astrophysics, University of California, Santa Cruz, CA 95064, USA}
\newcommand{\NSF}{National Science Foundation Graduate Research Fellow}
\newcommand{\ISEF}{ISEF \& De Gunzburg Fellowship}
\newcommand{\anuobs}{Mt Stromlo Observatory, The Research School of Astronomy and Astrophysics, Australian National University, ACT 2601, Australia}
\newcommand{\canterbury}{School of Physical and Chemical Sciences | Te Kura Matū, University of Canterbury, Private Bag 4800, Christchurch 8140, New Zealand}
\newcommand{\CIERA}{Center for Interdisciplinary Exploration and Research in Astrophysics (CIERA), Northwestern University, Evanston, IL 60208,USA}
\newcommand{\ICECSIC}{Institute of Space Sciences (ICE, CSIC), Campus UAB, Carrer de Can Magrans, s/n, E-08193 Barcelona, Spain}
\newcommand{\IEEC}{Institut d’Estudis Espacials de Catalunya (IEEC), E-08034 Barcelona, Spain}
\newcommand{\astrotd}{The ARC Centre of Excellence for All-Sky Astrophysics in 3 Dimension (ASTRO 3D), Australia}
\newcommand{\anucga}{Centre for Gravitational Astrophysics, College of Science, The Australian National University, ACT 2601, Australia}
\newcommand{\anu}{The Research School of Astronomy and Astrophysics, Australian
National University, ACT 2601, Australia}
\newcommand{\qub}{Astrophysics Research Centre, School of Mathematics and Physics, Queen's University Belfast, Belfast BT7 1NN, UK}
\newcommand{\ESO}{European Southern Observatory, Alonso de C\'ordova 3107, Casilla 19, Santiago, Chile}
\newcommand{\MIAMAS}{Millennium Institute of Astrophysics MAS, Nuncio Monsenor Sotero Sanz 100, Off. 104, Providencia, Santiago, Chile}
\newcommand{\warwick}{Department of Physics, University of Warwick, Gibbet Hill Road, Coventry CV4 7AL, UK}
\newcommand{\VT}{Department of Physics, Virginia Tech, 850 West Campus Drive, Blacksburg VA, 24061, USA}
\newcommand{\cifar}{CIFAR Azrieli Global Scholars program, CIFAR, Toronto, Canada}
\newcommand{\TelAviv}{The School of Physics and Astronomy, Tel Aviv University, Tel Aviv 69978, Isreal}
\newcommand{\birmingham}{Birmingham Institute for Gravitational Wave Astronomy and School of Physics and Astronomy, University of Birmingham, Birmingham B15 2TT, UK}
\newcommand{\OKC}{The Oskar Klein Centre, Department of Astronomy, Stockholm University, AlbaNova, SE-10691 Stockholm, Sweden}
\newcommand{\Uwarsaw}{Astronomical Observatory, University of Warsaw, Al. Ujazdowskie 4, 00-478 Warszawa, Poland}
\newcommand{\ctio}{Cerro Tololo Inter-American Observatory, NSF’s NOIRLab, Casilla 603, La Serena, Chile}
\newcommand{\lco}{Las Cumbres Observatory, 6740 Cortona Dr, Suite 102, Goleta,CA 93117-5575, USA}
\newcommand{\ucsb}{Department of Physics, University of California, Santa Barbara, CA 93106-9530, USA}
\newcommand{\oxf}{Department of Physics, University of Oxford, Oxford, OX1 3RH, UK}
\newcommand{\gemini}{Gemini Observatory, NSF's NOIRLab, 670 N. A'ohoku Place, Hilo, Hawai'i, 96720, USA}
\newcommand{\tcd}{School of Physics, Trinity College Dublin, The University of Dublin, Dublin, D02 PN40, Ireland}
\newcommand{\portsmouth}{Institute of Cosmology and Gravitation, University of Portsmouth, Burnaby Road, Portsmouth, PO1 3FX, UK}
\newcommand{\Carnegie}{The Observatories of the Carnegie Institution for Science, 813 Santa Barbara St., Pasadena, CA 91101, USA}
\newcommand{\tapir}{TAPIR, Walter Burke Institute for Theoretical Physics, 350-17, Caltech, Pasadena, CA 91125, USA}
\newcommand{\PSU}{Department of Astronomy \& Astrophysics, The Pennsylvania State University, University Park, PA 16802, USA}
\newcommand{\ICDS}{Institute for Computational \& Data Sciences, The Pennsylvania State University, University Park, PA 16802, USA}
\newcommand{\IGC}{Institute for Gravitation and the Cosmos, The Pennsylvania State University, University Park, PA 16802, USA}
\newcommand{\kavlicambridge}{Institute of Astronomy and Kavli Institute for Cosmology, Madingley Road, Cambridge, CB3 0HA, UK}
\newcommand{\dark}{DARK, Niels Bohr Institute, University of Copenhagen, Jagtvej 128, 2200 Copenhagen, Denmark}
\newcommand{\umel}{School of Physics, The University of Melbourne, VIC 3010, Australia}
\newcommand{\uhawaii}{Institute for Astronomy, University of Hawaii, 2680 Woodlawn Drive, Honolulu, HI 96822, USA}
\newcommand{\ucb}{Department of Astronomy and Astrophysics, University of California, Berkeley, CA 94720, USA}
\newcommand{\ncu}{Graduate Institute of Astronomy, National Central University, 300 Zhongda Road, Zhongli, Taoyuan 32001, Taiwan}
\newcommand{\fsu}{Department of Physics, Florida State University, 77 Chieftan Way, Tallahassee, FL 32306, USA}
\newcommand{\thacher}{The Thacher School, 5025 Thacher Rd., Ojai, CA 93023, USA}

\title{Flight of the Bumblebee: the Early Excess Flux of Type Ia Supernova 2023bee revealed by \textit{TESS}, \textit{Swift} and\\ Young Supernova Experiment Observations} 

\author[0000-0001-5233-6989]{Qinan~Wang}\blfootnote{Corresponding author: Qinan~Wang\\ \href{mailto:qwang75@jhu.edu}{qwang75@jhu.edu}}
\affiliation{\JHU}
\author[0000-0002-4410-5387]{Armin Rest}
\affil{\JHU}
\affil{\STScI}
\author[0000-0001-9494-179X]{Georgios~Dimitriadis}
\affiliation{\tcd}
\author[0000-0003-1724-2885]{Ryan~Ridden-Harper}
\affiliation{\canterbury}
\author[0000-0003-2445-3891]{Matthew~R.~Siebert}
\affiliation{\STScI}
\author[0000-0002-0629-8931]{Mark~Magee}
\affiliation{\warwick}

\author[0000-0002-4269-7999]{Charlotte~R.~Angus}
\affiliation{\dark}
\author[0000-0002-4449-9152]{Katie~Auchettl}
\affiliation{\UCSC}
\affiliation{\umel}
\affiliation{\astrotd}
\author[0000-0002-5680-4660]{Kyle~W.~Davis}
\affiliation{\UCSC}
\author[0000-0002-2445-5275]{Ryan~J.~Foley}
\affil{\UCSC}
\author[0000-0003-2238-1572]{Ori D. Fox}
\affiliation{\STScI}
\author[0000-0001-6395-6702]{Sebastian~Gomez}
\affiliation{\STScI}
\author[0000-0001-5754-4007]{Jacob~E.~Jencson}
\affiliation{\JHU}
\author[0000-0002-6230-0151]{David~O.~Jones}
\affiliation{\gemini}
\author[0000-0002-5740-7747]{Charles~D.~Kilpatrick}
\affiliation{\CIERA}
\author[0000-0002-2361-7201]{Justin~D.~R.~Pierel}
\affiliation{\STScI}
\author[0000-0001-6806-0673]{Anthony~L.~Piro}
\affiliation{\Carnegie}
\author[0000-0002-1633-6495]{Abigail~Polin}
\affiliation{\Carnegie} 
\affiliation{\tapir}
\author[0000-0003-3727-9167]{Collin~A.~Politsch}
\affiliation{\kavlicambridge}
\author[0000-0002-7559-315X]{César~Rojas-Bravo}
\affiliation{\UCSC}
\author[0000-0002-9301-5302]{Melissa~Shahbandeh}
\affiliation{\JHU}
\affiliation{\STScI}
\author[0000-0002-5814-4061]{V.~Ashley~Villar}
\affiliation{\PSU}
\affiliation{\ICDS}
\affiliation{\IGC}
\author[0000-0002-0632-8897]{Yossef Zenati}
\altaffiliation{\ISEF}
\affiliation{\JHU}

\author[0000-0002-5221-7557]{C.~Ashall}
\affiliation{\VT}
\author[0000-0001-6965-7789]{Kenneth~C.~Chambers}
\affiliation{\uhawaii}
\author[0000-0003-4263-2228]{David~A.~Coulter}
\affiliation{\UCSC}
\author[0000-0001-5486-2747]{Thomas~de~Boer}
\affiliation{\uhawaii}
\author[0000-0003-4522-9653]{Nico~DiLullo}
\affiliation{\thacher}
\author[0000-0002-8526-3963]{Christa~Gall}
\affiliation{\dark}
\author[0000-0003-1015-5367]{Hua~Gao}
\affiliation{\uhawaii}
\author[0000-0003-1039-2928]{Eric~Y.~Hsiao}
\affiliation{\fsu}
\author[0000-0003-1059-9603]{Mark~E.~Huber}
\affiliation{\uhawaii}
\author[0000-0001-9695-8472]{Luca~Izzo}
\affiliation{\dark}
\author[0000-0003-2720-8904]{Nandita~Khetan}
\affiliation{\dark}
\author[0000-0002-2249-0595]{Natalie~LeBaron}
\affiliation{\ucb}
\author[0000-0002-7965-2815]{Eugene~A.~Magnier}
\affiliation{\uhawaii}
\author[0000-0001-9846-4417]{Kaisey~S.~Mandel}
\affiliation{\kavlicambridge}
\author[0000-0002-1052-6749]{Peter~McGill}
\affiliation{\UCSC}
\author[0000-0003-2736-5977]{Hao-Yu~Miao}
\affiliation{\ncu}
\author[0000-0001-8415-6720]{Yen-Chen~Pan}
\affiliation{\ncu}
\author[0000-0003-0763-6004]{Catherine~P.~Stevens}
\affiliation{\VT}
\author[0000-0002-9486-818X]{Jonathan~J.~Swift}
\affiliation{\thacher}
\author[0000-0002-5748-4558]{Kirsty~Taggart}
\affiliation{\UCSC}
\author[0000-0001-7823-2627]{Grace~Yang}
\affiliation{\thacher}

\begin{abstract}

We present high-cadence ultraviolet through near-infrared observations of the Type Ia supernova (SN~Ia) 2023bee at $D=32\pm3$~Mpc, finding excess flux in the first days after explosion, particularly in our 10-minute cadence \textit{TESS} light curve and \textit{Swift} UV data.
Compared to a few other normal SNe~Ia with early excess flux, the excess flux in SN~2023bee is redder in the UV and less luminous. We present optical spectra of SN~2023bee, including two spectra during the period where the flux excess is dominant. At this time, the spectra are similar to those of other SNe~Ia but with weaker \ion{Si}{2}, \ion{C}{2} and \ion{Ca}{2} absorption lines, perhaps because the excess flux creates a stronger continuum.
We compare the data to several theoretical models on the origin of early excess flux in SNe~Ia. Interaction with either the companion star or close-in circumstellar material is expected to produce a faster evolution than observed.
Radioactive material in the outer layers of the ejecta, either from double detonation explosion or a $^{56}$Ni clump near the surface, can not fully reproduce the evolution either, likely due to the sensitivity of early UV observable to the treatment of the outer part of ejecta in simulation.
We conclude that no current model can adequately explain the full set of observations. We find that a relatively large fraction of nearby, bright SNe~Ia with high-cadence observations have some amount of excess flux within a few days of explosion. Considering potential asymmetric emission, the physical cause of this excess flux may be ubiquitous in normal SNe~Ia.


\end{abstract}



\section{Introduction}\label{sec:intro}

Type Ia supernovae (SNe Ia) are traditionally believed to be the thermonucelar explosion of a white dwarf (WD) star. Their standardizeable light curves at optical wavelengths \citep[e.g.][]{Phillips1993} serve as the foundation for measurements of the accelerating universe \citep{riess1998,perlmutter1999} and, therefore, for the entire $\Lambda$CDM model. Despite their important role, the exact SN~Ia explosion mechanism and progenitor system, as well as implications they may have on our cosmological interpretations, are not well understood \citep[see summary by][]{maoz2014observational}. Specifically, it is still not clear whether the progenitor systems are single- or double-degenerate (SD or DD), i.e., whether the companion is non-degenerate, like a main-sequence star \citep[][]{1973ApJ...186.1007W, Mazzali+07}, or also degenerate, like a WD \citep[][]{IbenTotukov84,Webbink1984, Fink+07, DanM+12, Moll_Woosley13,Pakmor+13,Liu+17, Shen_Kasen+18,Perets_Zenati+19}.

\cite{kasen2009seeing} originally suggested that the early SN~Ia light-curve could be a good way to distinguish between the SD and DD scenarios, sparking interest that has increased in recent years. In the canonical `expanding fireball' model, the early light curve of SNe~Ia is predicted to follow a $t^2$ law with negligible change in color under simplified assumptions \citep{arnett1982type}. In the SD scenario, however, \cite{kasen2009seeing} predicted that SN ejecta running into the binary companion and shocking it can produce excess flux on top of the power-law rise in the first few days after the explosion. Binary companions of different types and mass cause excess flux on time scales of days with varying brightness and colors directly after explosion.

Until recently, light curve ``bumps'' have remained elusive. Many detailed studies show that a power-law $L\propto t^\alpha$ with index $\alpha\sim2$ can serve as a good approximation in optical bands for both well-sampled, individual targets (i.e., SN~2009ig \citep{09ig}, SN~2011fe \citep{nugent2011} and ASASSN-14lp \citep{Shappee2016}) and large statistical samples of normal SNe~Ia \citep[e.g., ][]{Riess1999, Aldering2000, Goldhaber2001, Strovink2007, Garg2007, hayden2010rise, olling2015, 2020ApJ...902...47M}.  Several peculiar thermonuclear SNe have distinct ``bumps'' or excess flux beyond a normal power-law rise \citep[with the distinction that a bump requires a local maximum;][]{Cao2015, Jiang2017, De2019, miller20, Burke2021, 2020hvf, 2021zny}, but clear bumps in normal SNe~Ia appear to be rare.

However with the increase in high-cadence, early-time monitoring efforts, there is an increasing number of otherwise normal SNe~Ia with excess flux detected immediately after explosion, including SNe~2017cbv \citep{2017cbv}, 2018oh \citep{dimitriadis2018k2, Li2018oh, shappee2019seeing}, 2018aoz \citep{2018aoz} and 2021aefx \citep{Ashall2022, 2021aefxHosseinzadeh}. In general, these ``bumps'' are blue and last for $2$-$5$ days. In a study of 115 SN~Ia observed with the Zwicky Transient Facility (ZTF), \cite{Deckers2022} finds that 6 SN~Ia show evidence of flux excess within a few days of the explosion, and they conclude that $18\pm11$\% of SN~Ia have excess flux with simulated efficiency. However, we note that the cadence and signal-to-noise of the ZTF light curves are not at the same level as for SN~2017cbv, SN~2018oh, and SN~2021aefx. 

With the observational results, other possible mechanisms for producing excess flux have been suggested. For example, the shallow distribution of Ni$^{56}$ due to mixing can also change the shape of the early light curve of SNe~Ia \citep{PiroNakar2013, 2020magee1}. In the sub-Chandrasekhar mass (sub-$M_{ch}$) double-detonation (DDet) models, the first detonation in the helium shell produces a significant amount of radioactive isotopes in ashes, and might create an excess in early light curves \citep{Shen_Kasen+18,Polin2019,Perets_Zenati+19}. The characteristics of the early excess predicted by these various models vary in many respects, such as duration, amplitude, light curve shape, color evolution and rate. Thus, obtaining early fast-cadence observations in multiple bands is important to distinguish between the models.

To discover and monitor these rapid early signatures with high precision, continuous high-cadence photometric monitoring is necessary. The \textit{Kepler Space Telescope} \citep[\textit{Kepler};][]{haas2010kepler} and the \textit{Transiting Exoplanet Survey Satellite} \citep[\tess\,;][]{ricker2014} are designed with a large field-of-view (FOV) and cadence as high as 30 to 10 minutes, making them superb instruments not only for discovering exoplanets, but also for capturing light curves of extragalactic transients with exquisite cadence. \cite{olling2015} discovered 3 photometrically classified SNe Ia in the \kepler\ Prime mission, and found no evidence for flux excess in the early light curve.

Similar work was continued in K2, the successor of the \Kepler\ prime mission \citep{2014PASP..126..398H,rest2018fast,Ridden2019,ArmstrongTucker2021}. 
SN~2018agk shows a smooth power-law rise without early excess \citep{2018agk}. SN~2018oh, however, shows a prominent early excess within the first $\sim5$ days after the time of the explosion \citep{dimitriadis2018k2, Li2018oh, shappee2019seeing}. The morphology of this \kepler\ light curve is not constraining; it can be well-fit by the SD companion interaction model, the shallow $^{56}$Ni distribution model, and the sub-$M_{ch}$ DDet model. On the other hand, SN~2017cbv \citep{2017cbv,Sand2018} and SN~2021aefx \citep{Ashall2022, 2021aefxHosseinzadeh, 2021aefxNi}, covered by Swift and ground-based survey, also show early UV excess, but the fluxes are significantly weaker than those predicted by the SD companion interaction model. In other words, no single model can simultaneously explain both the early and late-time observations of SN~2018oh,  SN~2017cbv, and SN~2021aefx so far, but there have been a limited number of well-sampled light-curves due to \Kepler's relatively small FOV.

\tess\,, the successor of \kepler\,, has $\sim 20$ times larger FOV, but is also shallower by $\sim2$ magnitudes. This still means that \tess\ has the capability of increasing the number of high-cadence light curves of extragalactic transients by an order of magnitude. Already, \cite{2021zny} has revealed the existence of a 1.5-day duration bump in the early \tess\ light curve of a super-Chandrasekhar (03fg-like) SN~Ia, SN~2021zny, which can be explained by the SN ejecta interacting with a H/He-poor circumstellar medium (CSM). Combined with multi-band observations, \cite{2021zny} further demonstrates that the progenitor of SN~2021zny is likely to be a double carbon/oxygen WDs system, in which the less-massive WD had been tidally disrupted during the merger event and created a large amount of CSM before the supernova explosion. \tess\ also enables statistical studies on the properties of SN~Ia early light curves with high precision. With a sample of 24 normal SNe~Ia in 6 sectors in the first half year of \tess, \cite{fausnaugh2019early} found 3 of them with nearly linear rise, although no evidence of an additional component on top of the power-law rise has been found.

In this paper, we present the latest \tess\ SN~Ia, SN~2023bee, which shows an early excess flux detected in multiple bands. SN~2023bee was discovered within $\lesssim2$ days after the explosion and has been closely followed by ground and space-based facilities. Additionally, \tess\ also observed the SN from about $\sim12$ days before the explosion throughout the rising phase, and provides an extraordinary light curve with $10-$min cadence, though there is a $\sim 2$ day gap around the time of first light during which the images are heavily polluted by scattered light. The complete spectroscopic and photometric coverage makes SN~2023bee one of the best-observed SNe~Ia at early times, enabling a detailed study of the excess flux and its implication for progenitor properties. 
\cite{Hosseinzadeh2023bee} presents a different set of data, including an early spectral time series and radio observations.
In \S~\ref{sec:data} we present the acquisition and reduction of our data. In \S~\ref{sec:analysis} we analyze the early photometry and spectra. We present the model fits and discuss the implications in \S~\ref{sec:model}. The conclusions are presented in \S~\ref{sec:conclusion}.

\begin{figure}[t!]
    \centering
    \includegraphics[width=0.45\textwidth]{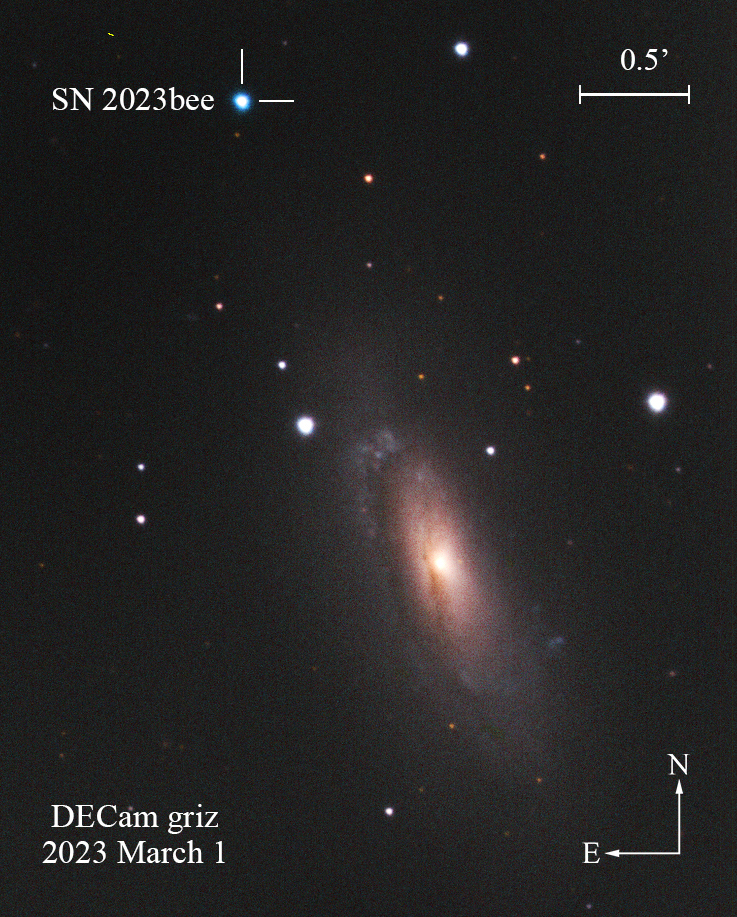}
    \caption{DECam composite $griz$ image stamp of SN~2023bee and its host NGC 2708 taken on 2023 Mar 1 UT, $\sim 10$ days after peak. The location of SN~2023bee is marked by white tick marks in the upper left corner. }
    \label{fig:finder}
\end{figure}

\section{Observations and Data Reduction} \label{sec:data}

SN~2023bee was discovered by the Distance Less Than 40 Mpc survey (DLT40) \citep{2017Yang,Tartaglia2018} on 2023 Feb 1 17:59:54.816 (MJD 59976.75) in the Clear band with an apparent magnitude of $17.26\pm0.04$ mag \citep{discovery}. SN~2023bee occurred at coordinates $\alpha=\rm 08^h56^m11.63^s$, $\delta= -03{\degr}19{\arcmin}32.06{\arcsec}$ (J2000.0) and was spectroscopically classified as a SN~Ia by \citet{classification2}. SN~2023bee is located at a distance of 93.97\arcsec\ from the center of its host galaxy NGC~2708, which is an intermediate spiral galaxy at $z = 0.0067 \pm 0.0005$ and distance modulus $\mu = 32.5\pm0.2$~mag \citep{2002ApJS..142..161P}.
Given the large separation between SN~2023bee and its host, the host extinction is likely to be negligible. Throughout this paper, we use the Milky Way extinction of $E(B-V)_{MW}=0.0145$ from the extinction map described in \cite{2011ApJ...737..103S}. 


\begin{figure*}[t!]
    \centering
    \hspace*{-0.2in}
    \includegraphics[width=\textwidth]{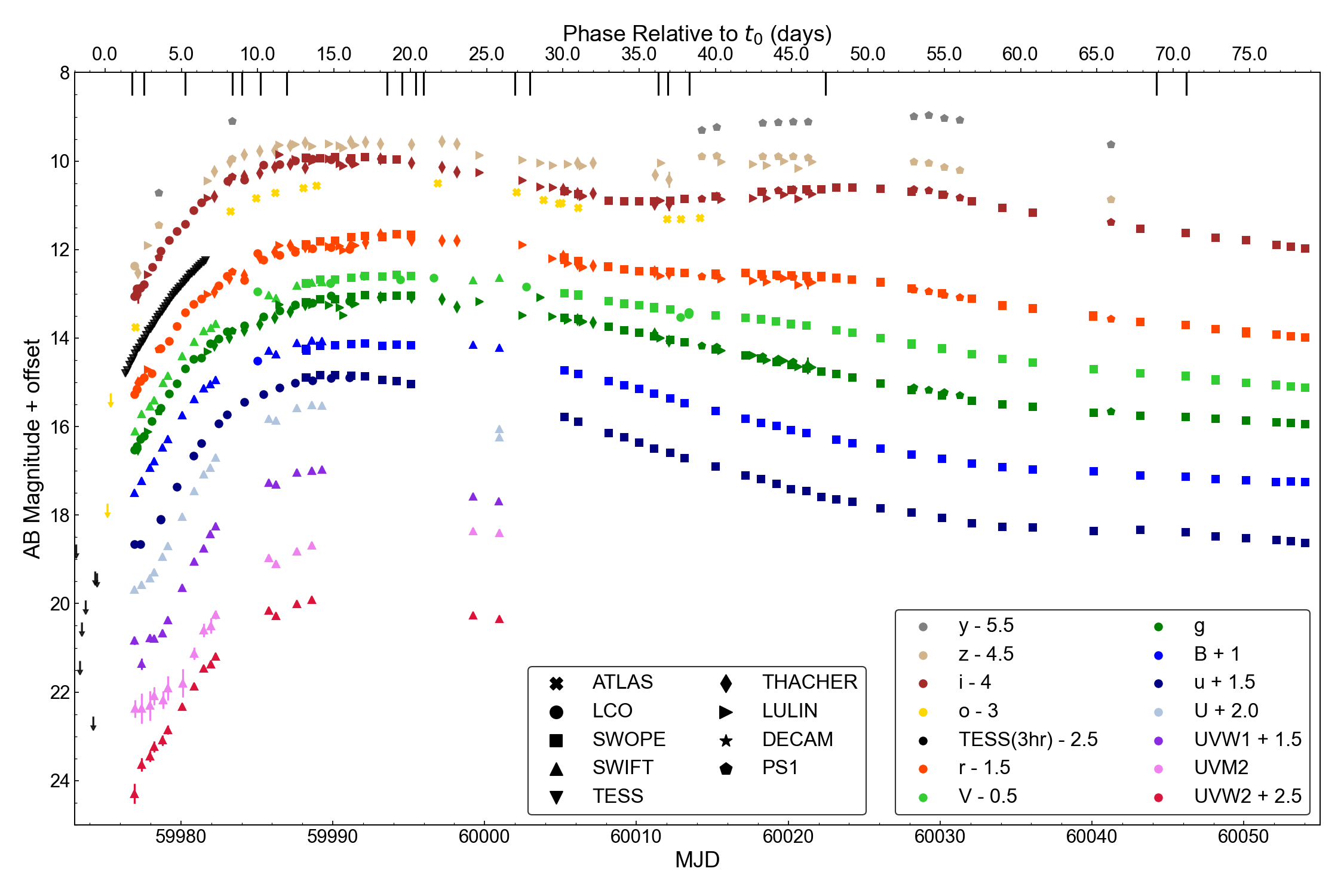}
    \caption{Multi-band light curves of SN~2023bee. The \tess\ data have been binned in 3-hour bins to increase the signal-to-noise ratio. The rest-frame phases relative to the inferred time of first light $t_0$ (see Section~\ref{sec:tess}) from \tess\ are labeled at the top. The down arrows mark the non-detections at the position of SN~2023bee in different bands and epochs. The times when optical spectra were taken are labeled as black ticks at the top. }
    \label{fig:full_lc}
\end{figure*}

\subsection{Photometry}

\tess\ observed the rise of 2023bee during sector 61, in CCD 1 of camera 1.  \tess\ features a broadband filter that covers the $r, i, z$ and $y$ bands with a wavelength range of 5802.57 to 11171.45~{\AA}. We create a $90\times90$~pixel$^2$ target pixel file, centered on 2023bee, from the calibrated TICA\footnote{doi:10.17909/t9-9j8c-7d30} \citep{TICA2020} full frame images (FFIs) with \texttt{TESScut}. The target pixel file was reduced with the standard \texttt{TESSreduce} pipeline, which accounts for image alignment, the scattered light background, and image artifacts to produce differenced images. We then conduct a secondary background subtraction by subtracting the median counts of each column from each pixel in that column, for every differenced image. 

The baseline flux is estimated by the median value between MJD 59969.2 and 59974 when the background is stable and no SN flux is present. Between MJD 59974.5 and 59976.3 the \tess\ measurements are significantly compromised by scattered light in the background and thus are excluded from further analysis. The complete \tess\ light curve along with details of the method and criteria we used to determine the compromised MJD range are described in \hyperref[appendix]{Appendix}.

We calibrate the \tess\ counts of SN~2023bee to physical AB magnitudes with the spectrum taken on MJD 59977.56 (see Fig.~\ref{fig:spec}). This spectrum is first calibrated using the LCO $g,r$, and $i$ measurements taken at the same night. Then we calculate a synthetic \tess\ magnitude, which we compare to the instrumental magnitude of the \tess\ measurement at the same time to compute the zeropoint. For this process, we use the \texttt{pyphot} package \citep{fouesneaumorgan_2022} and the bandpasses available from the Spanish Virtual Observatory (SVO) \citep[][]{2012ivoa.rept.1015R,2020sea..confE.182R}. We find the zero-point of \tess\ to be $zp_{TESS}=26.16\pm0.02$~mag.




We also observed SN\,2023bee in $griz$ with DECam at the CTIO 4~m Blanco telescope \citep{decam2008, 2015AJ....150..150F} and the PanSTARRS1 (PS1) telescope \citep{2016arXiv161205560C} as part of the Young Supernova Experiment \citep{Jones2021,Aleo2022}. Standard reductions for the DECam and PS1 images are performed by the NOIRLab community pipeline \citep{valdes_decam_2014} and the PS1 image Image Processing Pipeline (IPP), respectively \citep{2020ApJS..251....3M,2020ApJS..251....5M,2020ApJS..251....6M,2020ApJS..251....4W}. 
These images are then taken as input to the \texttt{photpipe} pipeline \citep{rest_testing_2005,rest_cosmological_2013}, which re-determines the zero points by comparing DoPHOT PSF photometry from each image to the Pan-STARRS DR1 catalog \citep{PS1catalogue}, convolves and subtracts a template image from the survey image, and performs forced photometry on the resulting difference images to create SN light curves. Fig.~\ref{fig:finder} shows a DECam $griz$ color image taken on 2023 Mar 1 UT. 

We observed SN\,2023bee with the Las Cumbres Observatory (LCO) 1\,m telescope network in $uBVgriz$ bands, the 1\,m telescope at Lulin Observatory using the Lulin Compact Imager, and with the Thacher 0.7\,m telescope in Ojai, CA from 2 February to 11 March 2023 in the $griz$ bands \citep{Swift2022}. 
Using the {\tt photpipe} imaging and reduction pipeline \citep{rest_testing_2005,rest_cosmological_2013}, we performed bad-pixel masking, reprojecting the data to a common pixel scale and pointing center using {\tt SWarp} \citep{swarp}, photometry with {\tt DoPhot} \citep{schechter_dophot_1993}, and photometric calibration using the Pan-STARRS 3$\pi$ \citep{PS1catalogue} and SkyMapper photometric catalogs \citep{onken2019}.  The final  photometry of SN\,2023bee was obtained by performing forced point-spread function (PSF) photometry at the average position of the source across all of our images.

Additionally, we observed SN 2023bee in optical $uBVgri$ bands with the Swope 1\,m optical telescope located at Las Campanas Observatory, Chile, as part of the Precision Observations of Supernova Explosions (POISE) \citep{POISE}. As described in \cite{kilpatrick2018}, all image processing and optical
photometry was performed using photpipe \citep{rest_testing_2005},
flat-fielding, image stitching, and photometric calibration. $BgVri$ photometry were calibrated using standard sources from the Pan-STARRS
DR1 catalog \citep{PS1catalogue}, while the u-band
data were calibrated using SkyMapper $u$-band standards \citep{onken2019}, transformed into the Swope natural system \citep{Krisciunas} with the Supercal
method \citep{Scolnic_supercal}. 

In addition, the {\it Neil Gehrels Swift Observatory} ({\it Swift}) observed SN\,2023bee from 1--26 February 2023.  We downloaded all processed Ultraviolet Optical Telescope (UVOT) data from NASA/HEASARC and performed forced aperture photometry at the location of SN\,2023bee using methods in {\tt heasoft} \citep[v6.28;][]{heasoft} and calibrated using the latest {\it Swift}/UVOT sensitivity files.

SN~2023bee was also observed by ATLAS, a project using four $0.5\,{\rm m}$ telescope systems installed on Haleakala (Hawaii), Mauna Loa (Hawaii), Las Campanas (Chile), and Sutherland (South Africa) to discover and monitor solar system objects. ATLAS observes in cyan (\textit{c}) and orange (\textit{o}) filters \citep{tonry2018atlas}. The ATLAS images are processed as described in \cite{tonry2018atlas}, and then photometrically and astrometrically calibrated using the RefCat2 catalog \citep{tonry18ref}. Template generation, image subtraction procedures, and photometric measurements are carried out following \cite{smith20}. We obtain forced photometry using the ATLAS forced photometry server \citep{Shingles21}. The forced photometry light curve is then cleaned up and the average flux for each night is calculated using ATClean \cite{Rest21,Rest23}.







\subsection{Spectra}

\begin{figure}[t!]
    \centering
    \vspace*{0.1in}
    \hspace*{-0.3in}
    \includegraphics[width=0.48\textwidth]{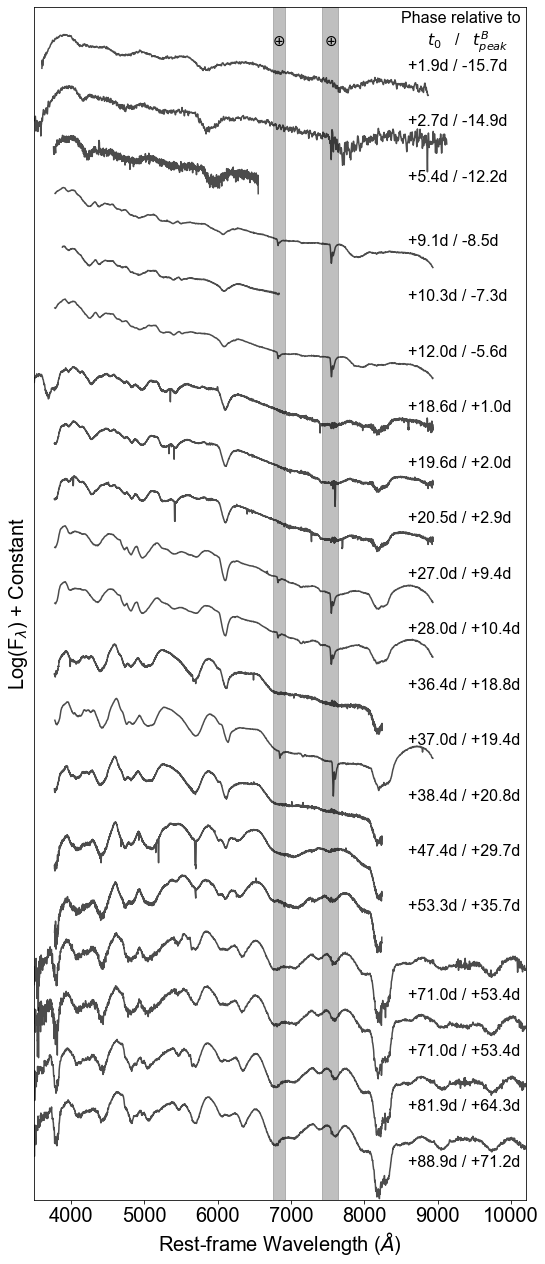}
    \caption{Optical spectra series of SN~2023bee. Phases relative to $t_0$ and $t_{peak}^{B}$ are labeled above each spectra. The telluric lines has been marked with the Earth ($\oplus$) symbol. All the spectra have been normalized and shifted for clarity.}
    \label{fig:spec_series}
\end{figure}

We collect 20 optical spectra from various sources, which include two publicly available spectra from the Transient Name Server \citep{classification1, classification2} obtained with the Lijiang 2.4m telescope and 2-meter Faulkes Telescope South (FTS) at Siding Spring Observatory (SSO) at early time, 7 spectra from Wide-Field Spectrograph (WiFeS) on the SSO 2.3m telescope, 6 spectra with the Alhambra Faint Object Spectrograph and Camera (ALFOSC) on the Nordic Optical Telescope (NOT) at La Palma, 3 spectra with the Kast spectrograph on the Lick 3\,m (Shane) telescope, 1 spectrum with the Kitt Peak Ohio State Multi-Object Spectrograph (KOSMOS, \citealt{KOSMOS}) on the Astrophysical Research Consortium (ARC) 3.5-meter Telescope at Apache Point Observatory (APO), and 1 spectrum with the Goodman spectrograph on the 4.1-meter Southern Astrophysical Research (SOAR) Telescope. 

The ALFOSC spectra were taken using grism 4 and a 1.0" slit, aligned along the parallactic angle, under clear observing conditions and good seeing ($<$1.5"). The spectra were reduced with a custom pipeline running standard pyraf procedures. 
The Kast and Goodman spectra were reduced through the {\tt UCSC Spectral Reduction Pipeline}\footnote{\url{https://github.com/msiebert1/UCSC\_spectral\_pipeline}} \citep{Siebert2020}, a custom data-reduction pipeline based on procedures outlined by \citet{Foley03}, \citet{Silverman2012}, and references therein. 
The WiFeS spectra were taken using a RT-560 beam splitter, a B3000 and R3000 diffraction gratings and Y=2 binning read out corresponding to a $1\times1$ arcsec spaxel. Each observation was reduced using PyWiFeS \citep{PYWIFES} producing a 3D cube file for each grating that has had bad pixels and cosmic rays removed. Spectra were extracted using QFitsView \footnote{\url{https://www.mpe.mpg.de/~ott/QFitsView/}} and a similar aperture to the seeing on the night (average seeing of $\sim$ 2"), while for background subtraction we extract a part of the sky that is isolated from the source. The KOSMOS spectra were reduced through the standard {\tt KOSMOS}\footnote{\url{https://github.com/jradavenport/pykosmos}} pipeline.
The spectroscopic time series is shown in Fig~\ref{fig:spec_series} and information of the spectra are listed in Table~\ref{tab:spec} in Appendix. A comparison to spectra from other SNe at similar phases is shown in Fig.~\ref{fig:spec}




\begin{table}
\centering
 \begin{tabular}{c c c} 
 \hline
 Parameters & SALT3 & Hsiao \\ [0.3ex] 
 \hline\hline
 $t_{0}/t^{B}_{peak}$ (MJD) &  $59973.796\pm0.006$ & $59992.354\pm0.005$ \\
  $x_0$ & $0.1017\pm0.0001$ & -   \\
  $x_1$ &$1.399\pm0.009$& -\\
  $c$& $-0.0902\pm0.0008$ & -\\
  Amplitude & - &$5.603\pm0.003\times10^{-6}$\\
\hline
\end{tabular}
\caption{Best-fit results for SALT3 and \cite{Hsiao07} models with \texttt{SNCosmo}. Note that SALT3 uses time of first light $t_0$ while Hsiao07 model uses time of B-band peak $t^{B}_{peak}$.}
\label{tab:lcfit}
\end{table}
\section{Analysis} \label{sec:analysis}

\subsection{Photometric Properties}\label{sec:salt}



In this section, we analyze the observed light curve of SN~2023bee and compare it to various models.
First, we fit the multiband light curves to the SALT3 \citep{salt3kenworthy, salt3pierel} model with \texttt{SNCosmo} \citep{sncosmo}, excluding the UV and \tess\ data due to its poor coverage in these wavelength ranges. The best-fit results for these two models are shown in Table~\ref{tab:lcfit}. The $B$-band peak is estimated to be $t_{peak}^{B} = 59992.58$ MJD with $m_{peak}^{B} = 13.041\pm0.001$ mag, corresponding to a $M_{peak}^{B} = -19.6\pm0.2$ mag. This corresponds to a $\Delta m_{15}$(B)$=0.788\pm0.002$ mag. Combining these properties, SN~2023bee is in the subclass of relatively slow and luminous SNe~Ia that are still considered normal SN~Ia \citep{Phillips1993,Hicken2009}, similar to SN~2021aefx \citep{2021aefxHosseinzadeh, Ashall2022}. 

We also fit the multi-band light curves using the spectral template model of \cite{Hsiao07}, which offers better coverage in the near-infrared (NIR) and allows us to include the exquisitely sampled \tess\ data in the fit.
The inferred time of the peak is $t_{peak}^{TESS} = 59991.8$ MJD with $m_{peak}^{TESS} = 13.68$~mag, and we will adopt these values in the later analysis.

\begin{figure}[t!]
    \centering
    \hspace*{-0.3in}
    \includegraphics[width=0.55\textwidth]{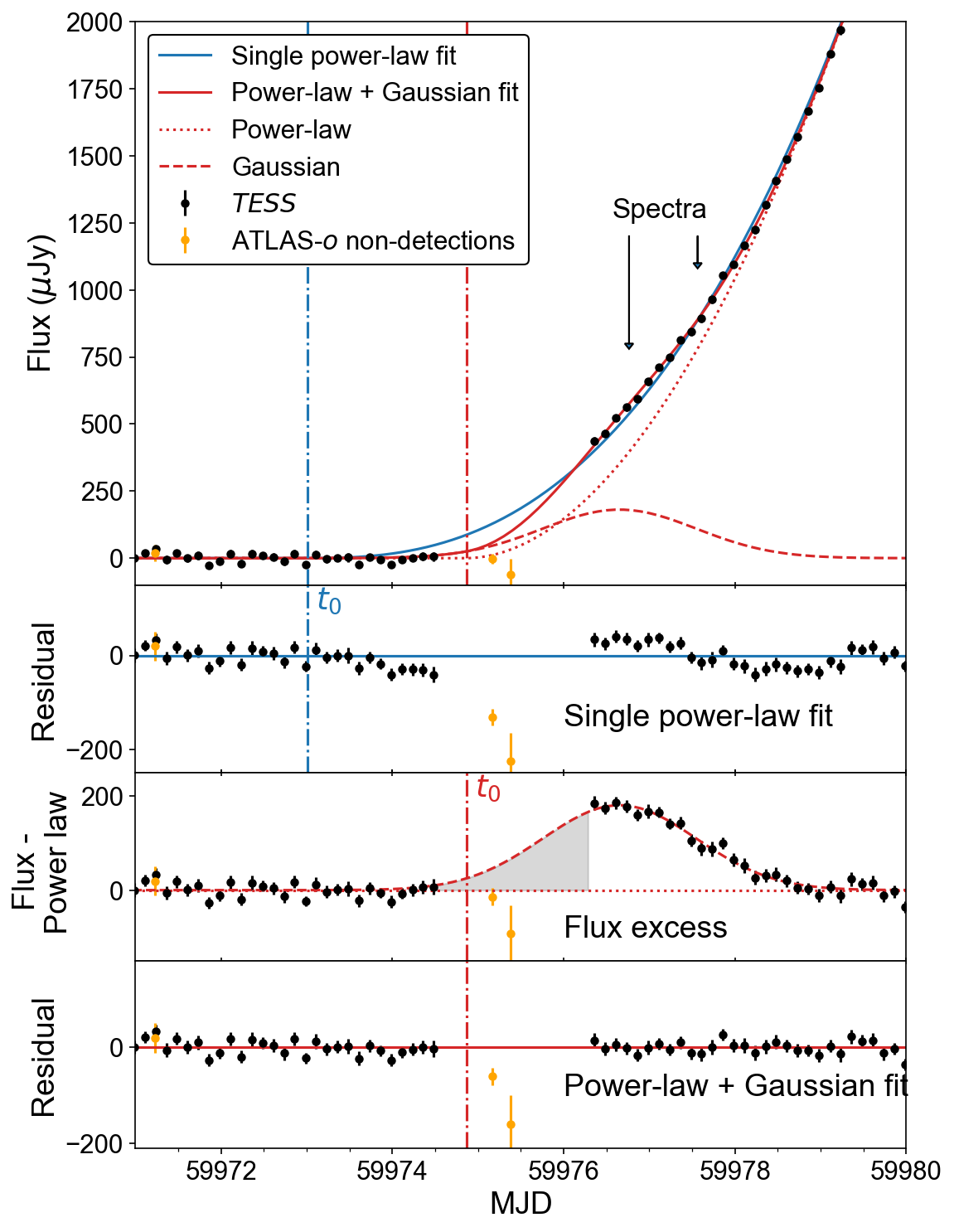}
    \caption{\textbf{Top}: the early \tess\ light curve of SN~2023bee fitted to a single power-law (blue solid) and power-law plus Gaussian (red solid). The power law and Gaussian component are plotted as dotted and dashed lines separately for clarity. The inferred $t_0$ of each power-law is plotted as a vertical line in the corresponding color. Down arrows denote the times when early spectra are obtained. \textbf{Middle}: residuals with regard to the single power-law fit. \textbf{Bottom panels:} the excess as flux subtracted by the power-law component in the power-law + Gaussian fit, and residuals with regard to the complete fit. The atlas-$o$ band non-detections are also included for comparison. In the shadowed region where the flux excess starts to rise, the Gaussian component is less constrained due to the lack of data, and thus may not correctly reflect the light curve during this period. 
    }
    \label{fig:pwvsgaussian}
\end{figure}

\begin{figure}[t!]
    \centering
    \hspace*{-0.2in}
    \includegraphics[width=0.5\textwidth]{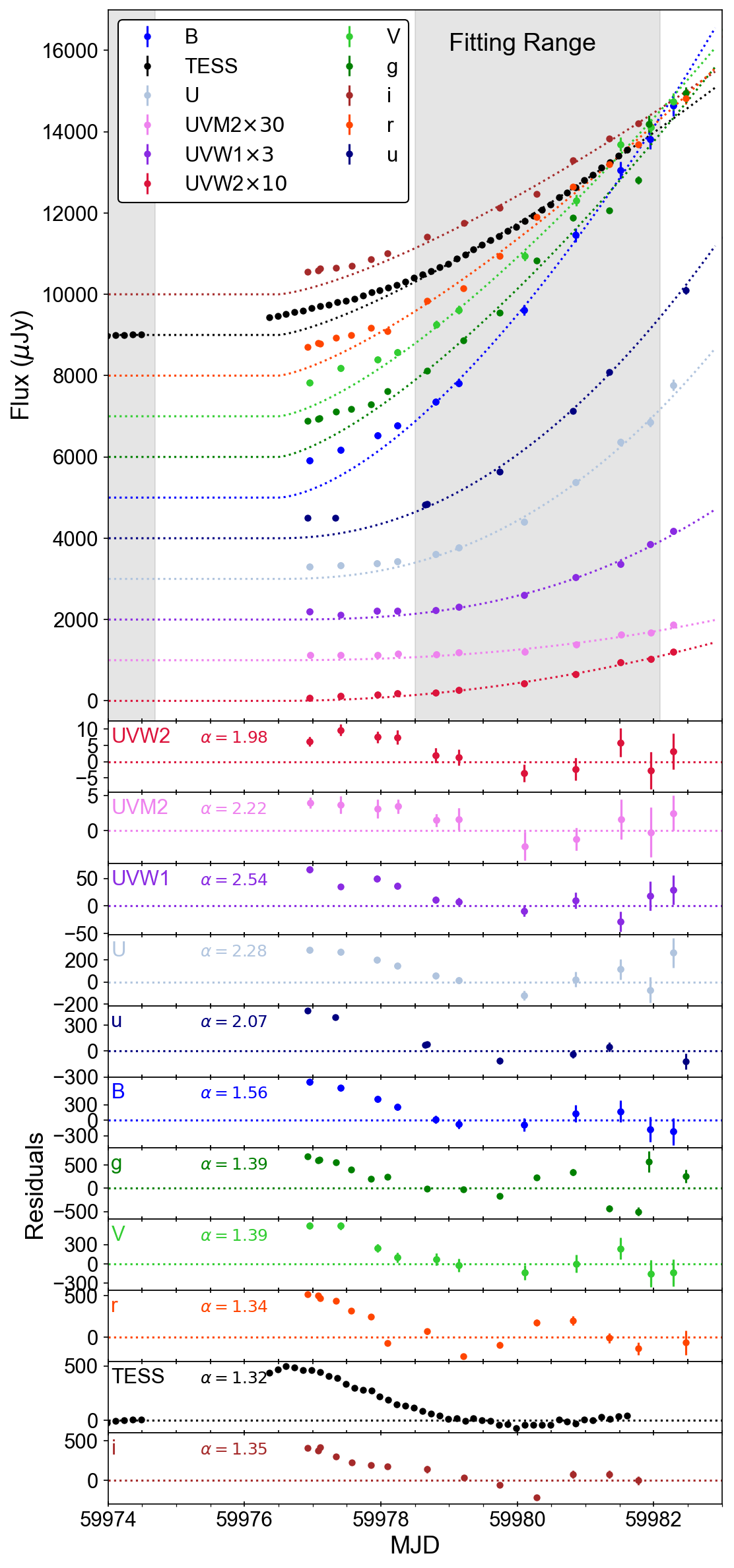}
    \caption{Partial power-law fit of the multi-band light curve of SN~2023bee and the residuals in different bands. The fitting range is marked as the grey region in the top panel. 
    The best-fit power-law index is listed in individual residual plots. 
    For better visualization, the data in the UV bands are rescaled as marked in the legend. 
    }
    \label{fig:full_lc_pwfit}
\end{figure}

\subsection{Early rise and excess}\label{sec:tess}

As seen in Fig.~\ref{fig:full_lc}, the UV light curves from {\it UVM2} to $U/u$ bands have a relatively flat rise within the first $\sim3$ days, indicating the existence of an early excess. However, in optical bands, the signature of early excess is more subtle and cannot be easily distinguished from the power-law rise of the SN itself. Thus, we make use of a few different power-law fitting schemes to verify the existence of the excess and determine the timepoint when the first optical emission emerge, i.e. the time of first light $t_0$ in optical bands.

\tess\ covers SN~2023bee from the pre-explosion stage until MJD 59981.67, when the SN reaches $\sim$30\% of the peak flux as estimated based on the $r$ and $i$ bands. Due to the influence of scattered light in the background, the \tess\ light curve around the time of the explosion is severely polluted, making it impossible to identify the first real detection and constrain $t_0$ with high precision. 
However, there are non-detections in the $o$ band on MJD 59975.17 and 59975.37 with upper limits of $m_o>20.76$ and $18.26$~mag respectively. These non-detections give a strong constraint on $t_0$. The first detection in $o$ band appeared on MJD 59977.01 with $m_o=16.76\pm0.02$~mag. Due to the lack of more $o$-band during the first days after explosion, it is impossible to fit a power-law to the $o$-band data directly. 


To identify whether there is excess flux in addition to the the power-law rise, and constrain $t_0$, we
fit the \tess\ light curve with two models: a single power-law rise,
\begin{equation}
  f(t) = A_{pl}(t-t_0)^{\alpha},
\end{equation}
and a power-law rise with a Gaussian function near the time of first light,
\begin{equation}
    f(t) = A_{pl} (t-t_0)^\alpha + \frac{A_{G}}{\sigma\sqrt{2\pi}} e^{-(t-\mu)^2/2\sigma^2}.
\end{equation}  
In the two equations, $t_0$ denotes the time of first light, $A_{pl}$ and $\alpha$ denote the scale and the index of the power-law, and $A_G$, $\mu$ and $\sigma$ denote the scale, center, and width of the Gaussian component respectively.
We note that the second scenario is unphysical since this model will never have zero flux. Nevertheless, we examine it to understand the morphology of the early light curve with a simple few-parameter function.

We perform a least-square fit with \texttt{lmfit} package \citep{lmfit}. The fitting results are shown in Figure~\ref{fig:pwvsgaussian} and Table~\ref{tab:fitresult}. To avoid overfitting, the Bayesian Information Criteria (BIC) is used to judge the goodness-of-fit in order to balance the deviation of fit and the number of parameters used. 

There are a few factors that make the power-law + Gaussian fit more preferable. First of all, this double-component fit has significantly lower BIC. Secondly, while the deviations from a single power-law are barely within $3\sigma$ limit in residual space for a given single flux measurement, there is a noticeable `S'-shape during the first few days, a characteristic sign of excess flux that cannot be fit with a single power-law with high significance \citep[e.g. see][]{dimitriadis2018k2, 2021zny}.
In addition, we have $o$-band non-detections in the \tess\ gap (see orange symbols in Fig.~\ref{fig:pwvsgaussian}). We can expect that the atlas-$o$ light curve is very similar to the \tess\ light curve, since their wavelength ranges are similar. The atlas-$o$ non-detection on MJD 59975.17 deviates from the single power-law fit with $>$5$\sigma$, which is another strong indication that the single-power law fit is not correct and the inferred $t_0$ is significantly biased toward early times.

For the power-law + Gaussian fit, we obtain $t_0 = 59974.86\pm0.26$ and we adopt this value in the following analysis. Note that in Figure~\ref{fig:pwvsgaussian} the model flux is non-zero before $t_0$ due to the tail of the Gaussian profile. Such a bias can be caused by the fact that the excess flux is likely asymmetric and cannot be fully described by a simple Gaussian. In principle this bias can be minimized by using a skewed Gaussian as shown in \cite{dimitriadis2018k2}, but the lack of data at early times corresponding to the first half of the Gaussian profile makes it difficult to constrain this asymmetry. Furthermore, the influence of this Gaussian tail is negligible in the residuals
Therefore, the skewed Gaussian fit gives indeed a higher BIC and is thus less useful in this case.

To see whether the flux excess exists in other bands, we also fit part of the multi-band light curves with the same $t_0$ and different power-law indexes, excluding the earliest phases when the flux excess is still significant compared to the power-law component. To find the best fitting range, we adopt an iterative fitting approach similar to \cite{dimitriadis2018k2}, using a fitting window with variable starting and ending times. To ensure adequate data points in multiple bands are included, the fitting goes through the starting time in between the first detection until when the \tess\ flux reaches $\sim 15\%$ of the estimated peak, and the ending time is set to be in between MJD 59982 to 59983.5 when the extrapolated power-law rise of \tess\ light curve lies in between $30\%$ to $40\%$ of the estimated peak. The MJD range with lowest reduced-$\chi^2$ is from 59978.5-59982.1 and is thus selected for the power-law fit, marked as the grey region in Figure~\ref{fig:full_lc_pwfit} along with the best-fit results and residuals.  The excess flux is clearly visible in all bands from the UV to $i$ band.

The fitted power-law indices are between 1.35 and 1.6 in the optical, which is significantly smaller than the power-law indices close to 2 found in SN~Ia without excess flux \citep{hayden2010rise,olling2015}. However, it is likely that at least some flux from the excess contributes to the total flux in the fitting region, affecting the exact power-law index. When we simultaneously fit the excess flux and power-law rise using theoretical models for the excess (see Section \ref{sec:model}), the power-law indices for optical bands is closer to the typical value of $2$ \citep{hayden2010rise,olling2015}. 

\begin{table*}[t!]
\centering
 \begin{tabular}{c c c c c c c c c} 
 \hline
 model & $t_{0}$ (MJD) & $\alpha$ & $A_{pl} (\mu$Jy) & $A_{G} (\mu$Jy) & $\mu$ (MJD) & $\sigma$ (days) & reduced $\chi^2$ & BIC\\ [0.3ex] 
 \hline\hline
power-law & $59973.00\pm0.15$ & $2.59\pm0.06$ &$17.4\pm3.1$& - & - & - & $518.1$ & $503.8$\\
power-law + Gaussian & $59974.86\pm0.26$ & $1.93\pm0.09$  & $115\pm29$ & $400\pm100$ & $59976.65\pm0.08$ & $0.90\pm0.09$ & $205.5$ & $440.7$\\
\hline
\end{tabular}
\caption{Different power-law fitting results on \tess\ light curve of SN~2023bee. }
\label{tab:fitresult}
\end{table*}

\subsection{Comparison with Other SNe~Ia in \tess\ and \kepler}

\begin{figure}[t!]
    \centering
    \vspace*{-0.32in}
    \includegraphics[width=0.5\textwidth]{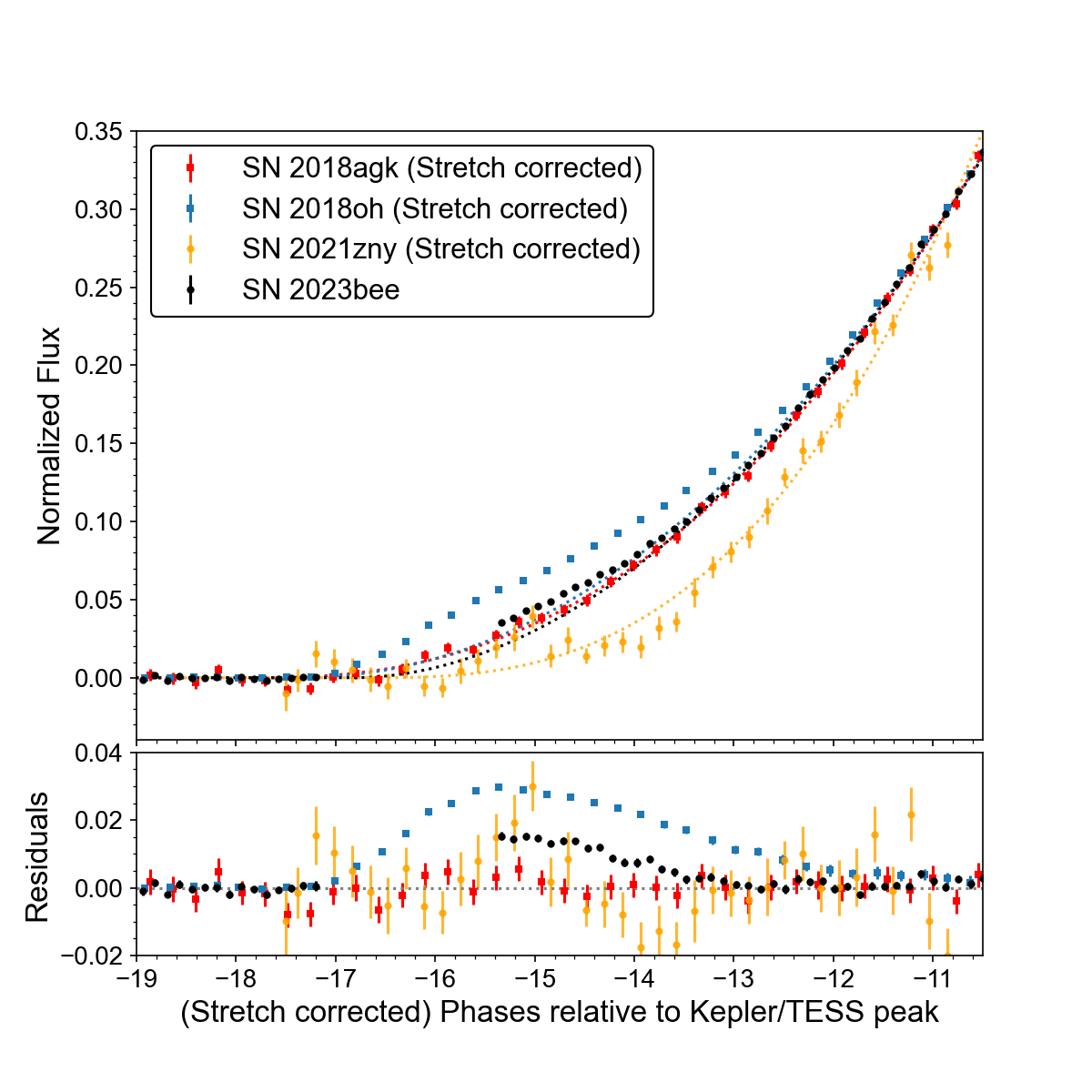}
    \caption{
    \textbf{Top:} Comparison of SN~2023bee (black) with other well-studied SNe~Ia with high cadence early light curves from \kepler\ and \tess, including SN~2018oh (blue), SN~2018agk (red) and SN~2021zny (yellow). The light curves of the comparison SNe~Ia have been `stretch-corrected' to match the rise of SN~2023bee, with regard to their rise time from $30$\% of the peak flux to the peak in the \kepler/\tess\ bands. Power-law fits to part of the light curves are also shown, for which the details are described in the main text. \textbf{Bottom}: Residuals relative to the power-law fits for these SNe~Ia in \kepler\ and \tess.}  
    \label{fig:ealy_time}
\end{figure}

We further compare the early \tess\ light curve of SN~2023bee to other well-studied SNe~Ia with high cadence light curve from \kepler\ and \tess. Our sample includes SN~2018oh, a normal SN~Ia with clear early excess in \kepler\ \citep{dimitriadis2018k2,shappee2019seeing}, SN~2018agk, a normal SN~Ia with a smooth power-law rise in \kepler\ \citep{2018agk}, and SN~2021zny, a super-$M_{ch}$ SN~Ia with short duration excess captured by \tess\ \citep{2021zny}. We adopt the partial power-law fits directly from the individual analyses.

We note that the sample shows a large spread in rise times.
SN~2023bee has a relatively short rise time $t_{\textrm{rise}} = 16.8$ days in the \tess\ band, while SN~2018oh and SN~2018agk have $t_{\textrm{rise}} = 18.2$ and $18.1$ days respectively, similar to the majority of normal SNe~Ia \citep[e.g., see][]{hayden2010rise, miller20}. SN~2021zny, in contrast, has a significantly longer rise time $t_{\textrm{rise}} \gtrsim 21$ days. We therefore `stretch correct' the comparison set of SNe~Ia with a stretch factor so that the other SNe fit best the rising arm of SN~2023bee from 30\% of the peak flux to the peak.
As shown in Figure~\ref{fig:ealy_time}, the power-law rise of SN~2023bee, SN~2018oh, and SN~2018agk align very well after `stretch correction,'. However, SN~2021zny is an outlier, and has a peculiar rise compared to the other SNe in this sample.
The excess in the \tess\ light curve of SN~2023bee has similar morphology to that of SN~2018oh in the \kepler\ band, though $\sim$50\% weaker and $\sim$1 day shorter. The excess of SN~2021zny, on the other hand, has a shorter duration and more abrupt evolution. It can also be seen from the residual plot that if the data for SN~2023bee had a similar S/N to that of SN~2018agk, the early excess would still be detectable. 

\subsection{Color Evolution}\label{section:color}

\begin{figure*}[t!]
    \centering
    \hspace*{-0.2in}
    \includegraphics[width=0.9\textwidth]{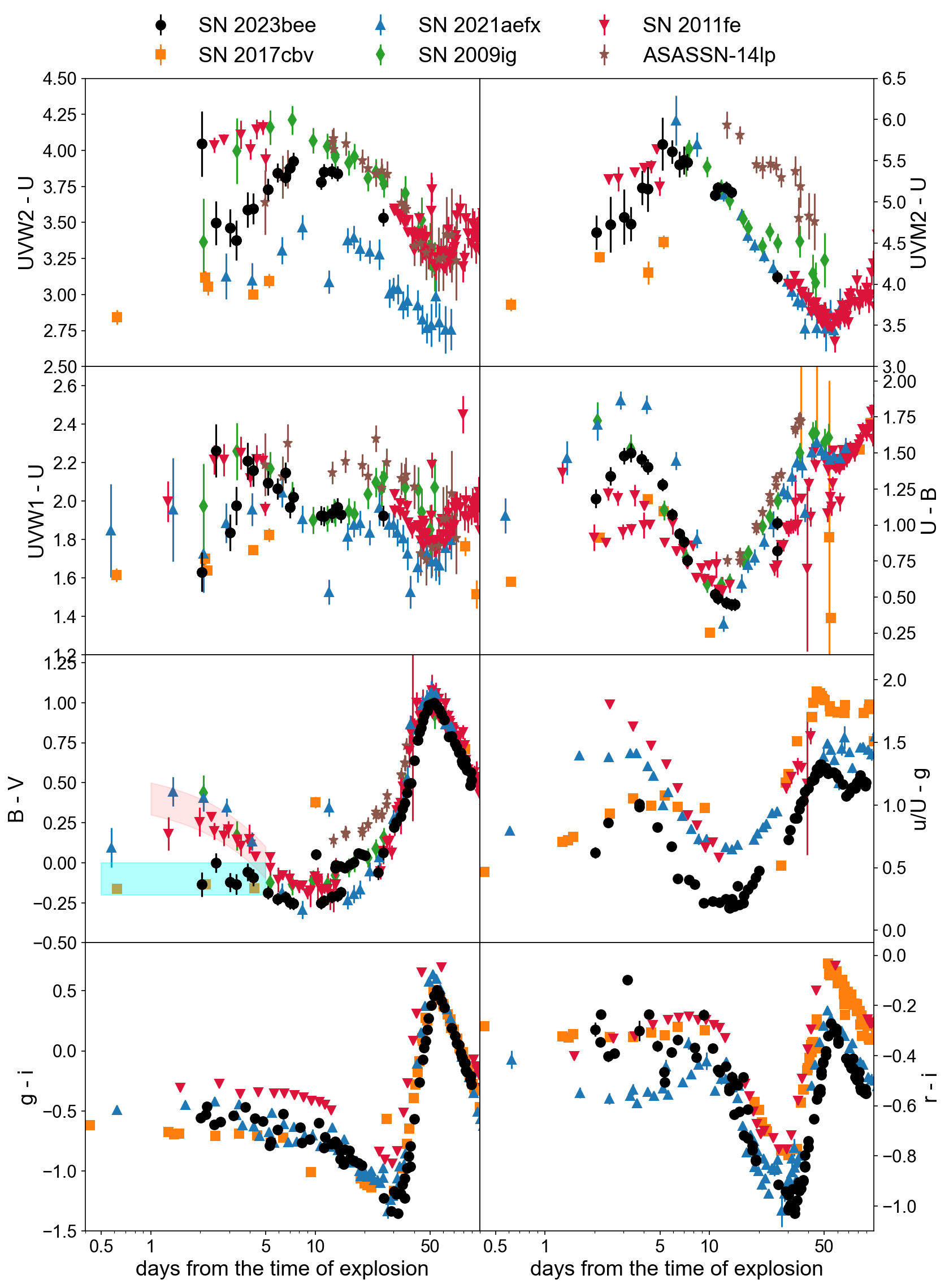}
    \caption{Extinction-corrected color evolution of SN~2023bee in multiple bands, in comparison with normal SNe~Ia (SN~2011fe, SN2009ig and ASASSN-14lp) and SN~Ia with bump in UV (SN~2017cbv, SN~2021aefx). All the magnitude have been converted into AB-magnitude system. In $U-B$, $B-V$ and all UV bands, the only data for SN~2023bee at early time are from \textit{Swift}. The cyan and red region in $B-V$ plot outline the evolution of `early blue' and `early red' subgroups as defined in \cite{Stritzinger2018}. The synthetic $ugri$ photometry of SN~2011fe are calculated from the photometrically calibrated spectra series from \cite{Pereira2013}.
    }
    \label{fig:color}
\end{figure*}



Figure~\ref{fig:color} shows the UV and optical color evolution of SN~2023bee at early phases, in comparison to other SNe~Ia with excesses flux such as SN~2017cbv \citep{2017cbv} and 2021aefx \citep{2021aefxHosseinzadeh, Ashall2022}, as well as other normal SNe~Ia without excess including SN~2009ig \citep{09ig}, SN~2011fe \citep{nugent2011} and ASASSN-14lp \citep{Shappee2016}. The color curves have been dereddened and converted to the AB magnitude system. Phases are relative to the time of first light estimated in the respective studies of each SN. 

Overall, in the UV bands, SN~2023bee has an intermediate color in between SNe~Ia with and without excesses, bluer than normal SNe~Ia and redder than SN~2017cbv and 2021aefx. 
\citet{Milne2013} separated SNe~Ia based on their UV colors at early times, and SN~2023bee would be NUV-blue based on this classification.  However, NUV-blue SNe tend to have lower ejecta velocities than NUV-red SNe \citep{Milne2015}, perhaps making SN~2023bee an outlier.
Similar to SN~2017cbv and SN~2021aefx, SN~2023bee also has a rapid trend in UVW1-$U$, $U-B$ and $u-g$ toward redder colors when the early excess is present. Afterward, the color evolution of SN~2023bee in UV resembles a normal SN~Ia as SN~2009ig and SN~2011fe, while SN~2017cbv and SN~2021aefx are significantly bluer around the same phases, while in some optical bands including $u/U-g$ and $r-i$, SN~2023bee is marginally bluer in later phases, similar to SN~2021aefx.

In the $B-V$ plot in Fig.~\ref{fig:color}, the regions correspond to the `early red' and `early blue' groups as defined by \cite{Stritzinger2018} have been highlighted, and SN~2023bee falls in the `blue' group where the $B-V$ color is relatively constant and blue, similar to SN~2017cbv. Given the observed colors across all of these bands, it is likely that there is a continuum from the `blue' to the `red' SNe~Ia rather than two distinct groups, in agreement with the conclusion of \citet{Bulla2020}.

\subsection{Spectroscopic Evolution}\label{section:comp}

\begin{figure}[t!]
    \centering
    \hspace*{-0.2in}
    \includegraphics[width=0.48\textwidth]{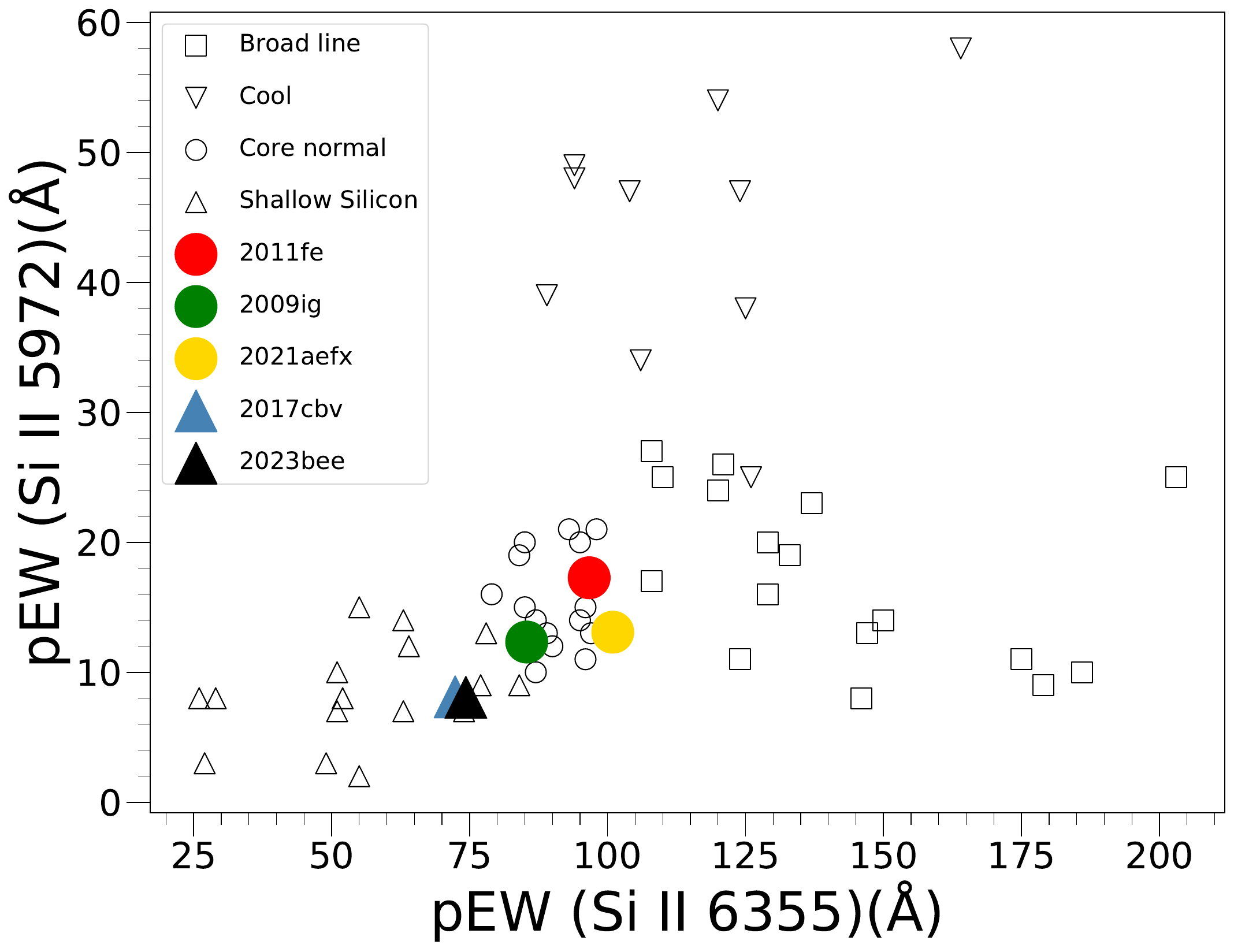}
    \caption{\ion{Si}{2} $\lambda$5972 and \ion{Si}{2} $\lambda$6355
  pseudo-equivalent-width (pEW) measurements for a sample of SN~Ia from \cite{Branch2009} (empty
  symbols) and for SNe~Ia of interest (filled symbols).  This
  parameter space defines subclasses of SN~Ia \citep{branch06}:
  ``shallow silicon,'' ``core normal,'' ``cool,'' and ``broad line''
  . 
  SN~2023bee falls within the ``shallow-silicon" subclass and is similar to another early excess SN~Ia, SN~2017cbv.}
    \label{fig:subclass_matt}
\end{figure}

\begin{figure}[t!]
    \centering
    \hspace*{-0.25in}
    \includegraphics[width=0.5\textwidth]{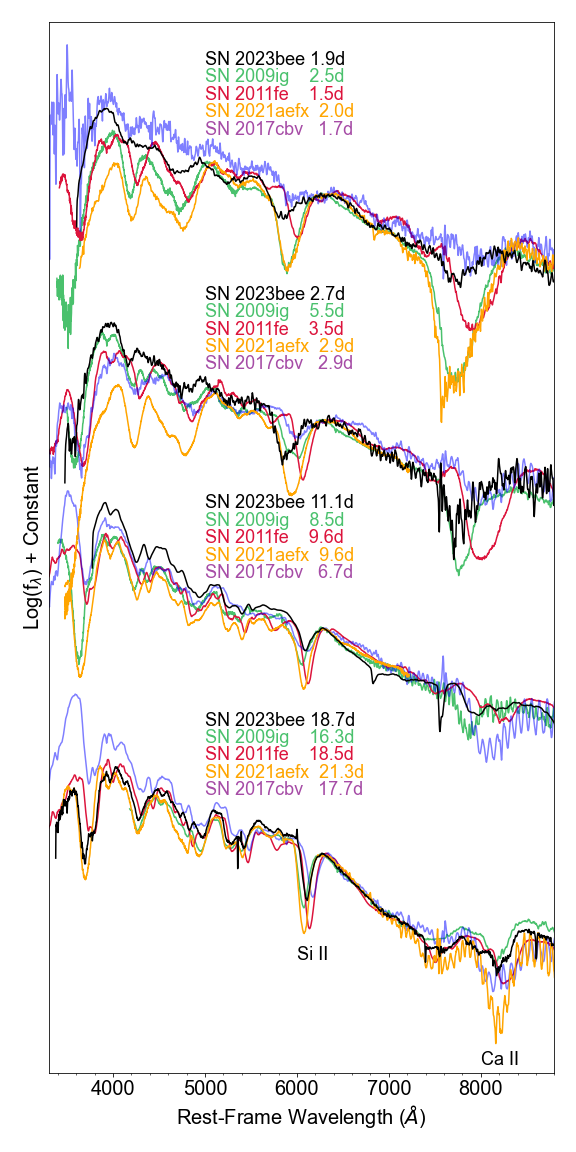}
    \caption{Comparison between spectra of SN~2023bee, SN~2009ig, SN~2011fe, SN~2017cbv and SN~2021aefx during the time of excess, rise, and around the peak. The phases relative to the inferred time of first light has been labeled around the spectra. The flux has been normalized to the continuum between $6250-6400$\AA. Note that SN~2023bee has shallow absorption features, similar to SN~2017cbv, but with significantly higher velocity.}
    \label{fig:spec}
\end{figure}

Two valuable spectra were obtained during the early flux excess.  In Figure~\ref{fig:spec}, we compare them with other SNe~Ia, including two SNe~Ia without an early flux excess (SN~2009ig, \citealt{09ig} and SN~2011fe, \citealt{2011feSpec}) and one SN~Ia with bright UV early excess (SN~2017cbv, \citealt{2017cbv} and SN~2021aefx, \citealt{2021aefxHosseinzadeh, Ashall2022}). The displayed spectra have all been normalized to the continuum between $6250-6400$~\AA\ for clarity.
At early times ($<$5.5d after the time of first light), SN~2023bee is most similar to SN~2017cbv with similar continuum and shallow absorption features, but with significantly higher velocity ($v_{\textrm{Si II}}\approx24,000$ km s$^{-1}$) in general. Each of these SNe displays peculiar absorption near the red wing of Ca H\&K, which has been interpreted as the result of high velocity \ion{Si}{2} $\lambda$4130 blending with Ca H\&K \citep{09ig}.

A notable difference between SN~2023bee and SN~2017cbv compared to other SNe~Ia without an early flux excess is their significantly weaker absorption features in spectra obtained within a few days of explosion. This difference is particularly striking in the \ion{Si}{2} $\lambda$6355, \ion{C}{2} $\lambda$6580 and \ion{Ca}{2} NIR triplet features with SNe~2017cbv and 2023bee having smaller equivalent widths than the comparison SNe. Since the same lines are present in all spectra with roughly the same line ratios, it is unlikely that there are significant differences in the abundance of intermediate-mass elements in the outer layers of these SNe. Contrastingly, only carbon and oxygen lines were detected in the earliest spectra of SN~2020esm, suggesting a significantly different composition than most SNe~Ia \citep{2020esm}. Instead, a difference in the equivalent widths may be the result of a stronger continuum corresponding to the excess flux at these times. On the other hand, the spectra of SN~2021aefx during the early excess phase are analogous to SN~2009ig and exhibit distinct line strength and SED compared to SN~2023bee and SN~2017cbv, potentially indicating the difference in the origin of early excess in optical bands. After the excess flux subsides, the equivalent widths of SNe~2023bee are closer to the comparison sample.

At maximum light, SN 2023bee appears to have the highest degree of similarity with SNe~2009ig and 2017cbv, having shallow \ion{Si}{2} features. The maximum-light spectrum is generally consistent with those of other slow-declining SNe ($\Delta m_{15}$(B) < 0.9 mag), having shallow \ion{Si}{2} $\lambda$5972 and a stronger peak on the blue side of Ca H\&K. We compare the \ion{Si}{2} absorption strength to other subclasses of SNe~Ia in Figure \ref{fig:subclass_matt}. SN~2023bee clearly falls within the boundary of the ``shallow-silicon" SN~Ia subclass identified in \citet{branch06}. Interestingly, SN~2017cbv, another early-excess SN~Ia had very similar \ion{Si}{2} absorption features to SN~2023bee at maximum light.
 

\section{modeling}\label{sec:model}

In this section, we discuss the different possible physical mechanisms that might explain the early excess flux, and the implications on the progenitor properties of SN~2023bee. 

\begin{figure*}[t!]
    \centering
    \hspace*{-0.2in}
    \includegraphics[width=\textwidth]{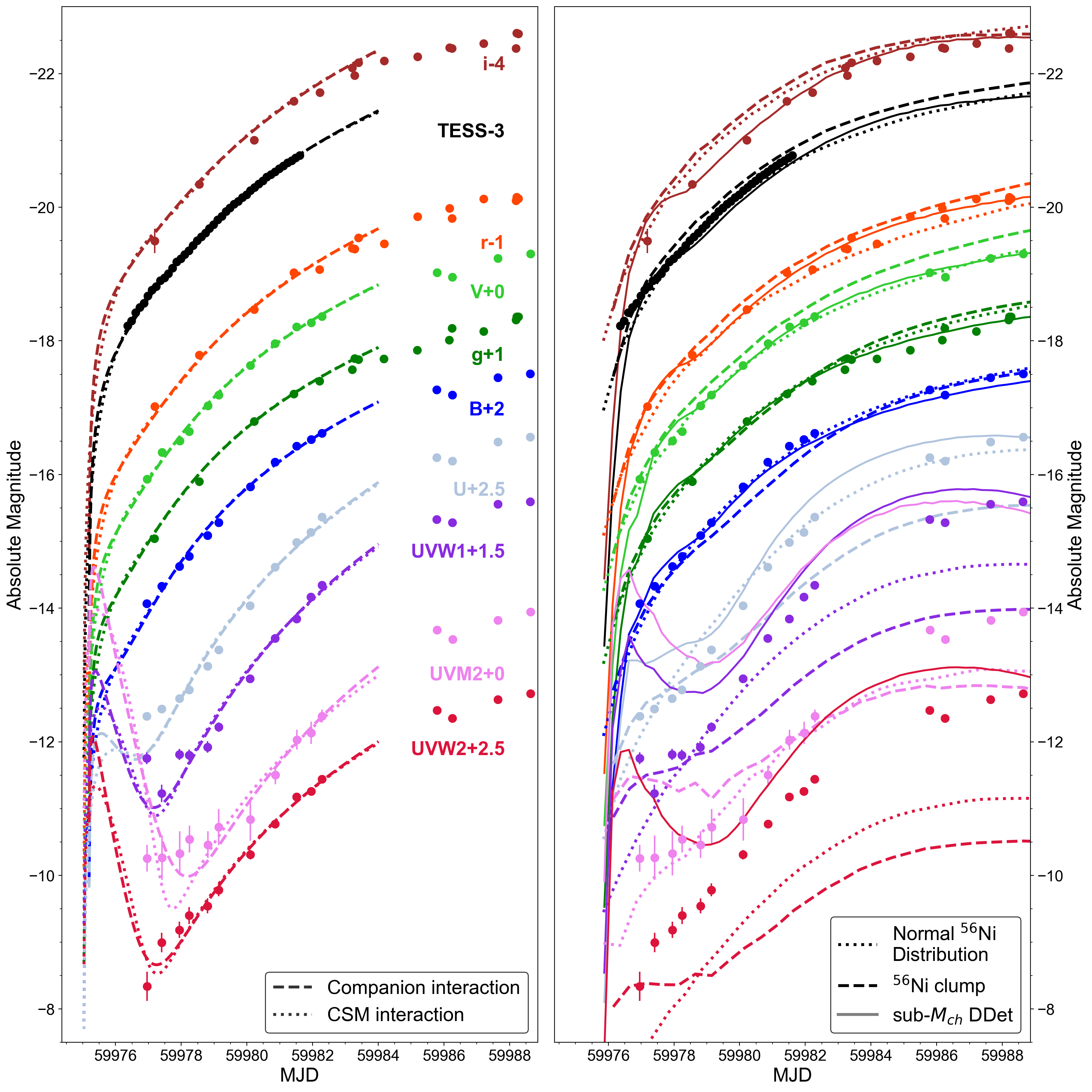}
    \caption{Multiband fit to the CSM-interaction and companion interaction models (left), and to the $^{56}$Ni mixing and sub-$M_{ch}$ double-detonation models (right). A model with normal $^{56}$Ni distribution from \citealt{2020magee1} is included as a fiducial model for comparison. Color for different bands are labeled below each light curve in the left panel. }
    \label{fig:modelfit2}
\end{figure*}

\subsection{Companion Interaction}

\cite{kasen2009seeing} illustrates that when the SN ejecta collide with a non-degenerate companion, a blue and luminous excess flux will arise in the first few days after the explosion. The luminosity, duration, and SED of the excess are dependent on the binary separation, ejecta velocity, and viewing angle. Following the same fitting scheme as described in \cite{2021zny}, we fit the multi-band light curves of SN~2023bee to this companion-interaction model, adopting an ejecta velocity $v_{\mathrm{ej}}=12,500\:\mathrm{km\:s^{-1}}$ as measured from the peak spectrum, and assuming the underlying SN light curve to be power-law with the same time of the explosion but different indices in different bands. The time of first light $t_0$ and binary separation $a$ are the free parameters in the excess model. Due to the severe degeneracy between the viewing angle and other parameters, we assume a viewing angle $\theta = 0^\circ$. 

The best-fit result is shown in the left panel of Fig.~\ref{fig:modelfit2} in comparison with the multiband light curve. While this model matches the rise of SN~2023bee relatively well, it has an early spike in the UV bands as a result of the high blackbody temperature, and thus fails to reproduce the relatively flat shape in multiple bands, especially for $U$, UVW1, and UVM2. One explanation for this discrepancy is that the models are not yet sophisticated enough to correctly predict the behavior in the bluer bands of the ejecta interactions with the companion star. For example, \cite{kasen2009seeing} assume a blackbody SED for the excess flux, but UV line blanketing if any could dramatically change the evolution of early flux. New, improved companion interaction models would significantly help with the interpretation.
Also, the power-law may not serve as a good approximation for individual SNe~Ia in certain bands, especially in UV where there are less the SNe~Ia with early coverage.
Improving the modeling of early SNe~Ia light curves in the different bands with new simulations and a larger data sample is necessary. 

The best-fit binary separation is $a = 2.27\pm0.14R_\odot$. Given the condition that the single-degenerate systems are believed to be undergoing Roche-lobe overflow, the companion size can also be correlated with the binary separation \citep{Eggleton1983}. Assuming a typical mass of companion to be $1-6M_\odot$ we can estimate that $R\sim0.8-1.15R_\odot$. These values are degenerate with the viewing angle so we are unable to give a conclusive answer on the companion parameters, but given the order of magnitude, this relatively small radius is in agreement with a main sequence star companion. This result is in agreement with the non-detection in the radio in \cite{Hosseinzadeh2023bee}, which indicates a low mass-loss rate and rules out most of the red-giant companions. The inferred $t_0 = 59975.18\pm0.02$ MJD is also marginally in agreement with our value inferred from the power-law + Gaussian fit. 


\subsection{CSM Interaction}


We also attempt to fit the multi-band light curves with a model similar to \cite{Ni2023ApJ}, \cite{2021zny}, and \cite{Srivastav2023ApJ}, in which the early flux excess is powered by confined CSM close to the explosion site, approximated as a spherically symmetric envelope of mass $M_{\mathrm{env}}$ and radius $R_{\mathrm{env}}$ (see \cite{Piro2015ApJ} for a complete presentation of the model). This model has successfully explained some relatively short and weak flux excesses, commonly seen in 03fg-like SNe (e.g. SN~2021zby), and possibly associated with double white dwarf merger events. Similar to our \cite{kasen2009seeing} model fit, we use a power law as the underlying SN light curve. As no hydrogen has been identified in the spectra of SN\,2023bee, we assume for the H-poor CSM an electron-scattering opacity of $\kappa=0.2\:\mathrm{cm^{2}\:g^{-1}}$, while for the ejecta mass and velocity, we use the canonical values of $M_{\mathrm{ej}}=1.4\:\mathrm{M_{\odot}}$ and $v_{\mathrm{ej}}=12,500\:\mathrm{km\:s^{-1}}$ as measured from the peak spectrum. Due to the lack of a robust estimate of the explosion date and the sparse coverage of the flux excess, we assume that the time of the onset of the CSM interaction coincides with the time of first light for each photometric band. 

The best-fit parameters are $M_{\mathrm{env}}=0.05\pm0.02\:\mathrm{M_{\odot}}$ and $R_{\mathrm{env}}=(6.5\pm1.4)\times10^{10}$ cm, and assuming a $\rho \sim r^{-3}$ density distribution for the envelope, we estimate $M_{\mathrm{CSM}}=0.3\pm0.1\:\mathrm{M_{\odot}}$, with $t_0 = 59975.02 \pm 0.06$ MJD. As can be seen from Fig.~\ref{fig:modelfit2}, similar to the companion interaction model, the CSM interaction model does not reproduce the relatively flat rise at early times. In fact, the CSM interaction model predicts an even more abrupt and shorter excess compared to the companion interaction model, making it less preferable in the case of SN~2023bee.


\subsection{Sub-Chandrasekhar Double-detonation Models}

In the sub--$\textrm{M}_{ch}$ DDet scenario, \cite{Polin2019} predict that the initial ignition of the thick He-shell on the surface will create radioactive material embedded in ashes in the outer layers of ejecta, which can cause early excess in the light curve. In general, a thicker He-shell will produce more radioactive isotopes and generate a stronger early excess. Due to the significant line blanketing in UV by these ashes, the early excess predicted by this DDet model tend to be red.

We tried fitting the data with all 39 1-dimensional models in the grid from \cite{Polin2019} using the least $\chi^2$ method to find the best match. Additionally we add two complementary parameters, the time of first light $t'_0$ and distance modulus $\mu'$, to account for the potential uncertainty in the phases and magnitudes. The range of $t'_0$ and $\mu'$ are set to be $\pm 2$ days around $t_0$ and $\pm 0.5$ mag around $\mu$. The best-fit model with the lowest $\chi^2$ is the one of a $1.1 M_\odot$ WD and He-shell mass $M_{shell} = 0.05 M_\odot$, with $t'_0 = 59975.75$ MJD and $\mu=32.35$ mag, and it is plotted as the solid line in the right panel of Fig.~\ref{fig:modelfit2}. 

Overall, sub--$\textrm{M}_{ch}$ DDet model has a reasonable match to the light curve of SN~2023bee in most optical bands, but from $U$ to UV bands this model severely overpredicts the flux for both SN rise and the early excess. Similar to the companion interaction model and CSM interaction model, it prefers a `spike'-like excess and fails to reproduce the relatively smooth rise of SN~2023bee. 
On the other hand, the constant $g-r$ color curve and a relatively weak absorption feature in spectra at the early time agree with the prediction of a thick He-layer with $M\gtrsim0.08M_\odot$ (see Fig.~6 and 8 in \citealp[][]{Polin2019}). Still, a thicker He-layer will produce a more prominent early `spike' in the UV, which can also be seen in the 12 models in \cite{Perets_Zenati+19}. Thus, it is difficult to reconcile the early color and spectral features with the flat UV rise of SN~2023bee in the sub$-\textrm{M}_{ch}$ DDet model. 
Another issue with sub$-\textrm{M}_{ch}$ DDet model is that the \ion{Si}{2}$\lambda$5972 velocity and relatively high peak luminosity of SN~2023bee seems to fall into the range of the $\textrm{M}_{ch}$ group.
We caution that the early UV observables will be highly sensitive to the treatment of the outer edge of the ejecta in simulations and modeling choices made in how to treat the surrounding medium. Better results with a flat rise in the UV can possibly be obtained by tracking the early outflow and unbound material during the first few orbital timescales in the Roche-Lobe Overflow stage until the merger in CO WD--HeCO WD DDet models \citep{Zenati+19, Perets_Zenati+19,Pakmor+21}, or by taking line-of-sight effect into account with multi-dimensional models \citep{Shen2021, Shen2021b}.




\subsection{$^{56}$Ni Clump}

We follow the method described by \cite{2020magee1} to find the best-matching $^{56}$Ni distribution model among their 1-dimensional model suite. In addition, we also include the $^{56}$Ni clump models presented by \cite{2020magee2} to determine whether they can provide a reasonable match to the early excess. Again, we allow for flexibility in the time of first light $t'_0$ and distance modulus $\mu'$. The best-matching $^{56}$Ni distribution and $^{56}$Ni clump models are EXP\_Ni0.8\_KE0.50\_P4.4 (no excess) with $t'_0=59975.3$ MJD and $\mu'=32.15$ mag, and SN2017cbv\_Ni0.04\_Mean1.350\_StdDev\_0.180 (with excess) with $t'_0=59975.4$ and $\mu'=32.6$ mag, respectively. Both models are shown in Fig.~\ref{fig:modelfit2}.

\par

As shown by Fig.~\ref{fig:modelfit2}, the $^{56}$Ni clump model does not reproduce the shape of the early excess and generally shows a more prominent bump than what is observed in SN~2023bee, particularly in the UV bands. We note that this model was designed around reproducing the light curve of SN~2017cbv and therefore it is unsurprising that it does not provide perfect agreement with SN~2023bee. Models designed specifically around SN~2023bee likely could provide improved agreement in the optical bands and around maximum light, however, a suppression of flux in the UV is a natural consequence of large $^{56}$Ni clumps in the outer ejecta as a result of significant line blanketing. Therefore while $^{56}$Ni clump models adapted for SN~2023bee may be able to better match the shape of the early excess, it is unlikely they would simultaneously match the UV observations. Similarly, the $^{56}$Ni distribution model shown in Fig.~\ref{fig:modelfit2} provides reasonable agreement in the optical, but cannot reproduce the shape of the excess, which is expected for these models. Again we find that the model does not match the UV observations. This likely results from an extended $^{56}$Ni distribution being preferentially selected in order to more closely match the excess at early times, which results in some UV line blanketing.

\par

Based on the overall agreement of the light curve and strong disagreement in the UV bands, we find that the early excess in SN~2023bee is unlikely to have resulted from surface $^{56}$Ni in a Chandrasekhar mass explosion. Further modeling exploring metallicity effects may provide improved agreement but is unlikely to overcome the significant discrepancies in the UV.

\par


\section{Discussion and Conclusion}\label{sec:conclusion}

In this paper, we presented early photometric and spectroscopic observations of SN~2023bee, including \textit{Swift} UV and $10-$min cadence \tess\ light curves starting $\sim2$ days after the time of first light.
SN~2023bee has a relatively short rise time ($\sim17$ days in \tess), high peak luminosity ($M^B_{peak}=-19.6$), and slow decline rate ($\Delta m_{15}(B)=0.788$). Most importantly, 
SN~2023bee shows clear evidence of an excess flux at early times detected in all bands, but most prominent in the UV.
The two early spectra of SN~2023bee, taken within $\sim3$ days after time of first light and at similar phases to the flux excess, show shallow high-velocity \ion{Si}{2} and \ion{Ca}{2} absorption features. These features are similar to those in SN~2017cbv, a Type Ia SN with an early excess flux, but distinctive from normal SNe~Ia that do not show an early excess. They are also different to SN~2021aefx, which has an early bump, but shows strong \ion{Si}{2} and \ion{Ca}{2} absorption features at similar phases. 

We use four different models to fit the early excess flux: companion interaction, CSM interaction, sub $M_{ch}$ DDet, and $M_{ch}$ $^{56}$Ni clump models. None of these models manage to accurately reproduce the light curves in all bands. All models except for the $^{56}$Ni clump model predict a sharp peak at the earliest phase in the UV, which is not observed in the early light curve of SN~2023bee. For both the $^{56}$Ni clump model and the sub-$M_{ch}$ DDet model, the overall relative scaling between different bands does not fit the observed light curve. Similar challenges in modeling their light curves have been experienced in other SNe~Ia with early UV excess such as SN~2017cbv \citep{2017cbv} and SN~2021aefx \citep{2021aefxHosseinzadeh, Ashall2022}.
The fact that none of the models explored here can adequately fit the observations over all passbands may have two explanations: either none of the physical mechanisms represent the true source of the excess and we must look for different explanations; or the current models are not sophisticated enough and/or do not explore a large enough parameter space to accurately reflect the complexity of the light curves. This is particularly true in the UV, where small changes are difficult to parameterize from initial conditions and, for example line blanketing, can lead to significant changes in the predicted spectra and photometry. More UV data, especially rapid UV spectroscopy from the \textit{Hubble Space Telescope} within the first few days after explosion, like the UV spectra obtained during the rapid rise of SN~IIP~2020fqv \citep{2020fqv}, would uniquely probe the physical models and explain the nature of the excess.


Currently, even exquisite high-cadence multiband early observations of early excess SNe~Ia such as the one presented here are unable to reliably distinguish between different progenitor scenarios. The nearby nature of these events makes it possible to observe them well into the nebular phase. Models that predict early flux excesses have distinct predictions for late time nebular emission. For example, narrow H/He features could be indicative of a single degenerate companion \citep{kasen2009seeing, Kollmeier2019, vallely2019,Prieto2020, Elias-Rosa2021}, as seen in SN~2019yvq where strong [\ion{Ca}{2}] emission could be the result of sub-$M_{ch}$ double detonation though the model prediction does not simultaneously match the early light curve\citep{Siebert2020, Polin2019,Burke2021, Tucker2021}, and [\ion{O}{1}] emission, like that seen in the ``02es-like" SN~2010lp and iPTF14atg maybe be produced in violent merger events \citep{Taubenberger13,Kromer16}. Furthermore, early flux excesses may occur at a higher rate in super-Chandrasekhar ``03fg-like" SNe~Ia \citep{Jiang18}. These tend to have broad [\ion{O}{1}] and sharp [\ion{Ca}{2}] emission \citep{Taubenberger17, 2021zny}. Given the remarkable photometric and spectroscopic similarity to SN~2017cbv and SN~2018oh, we may expect SN~2023bee to have similar behavior at late times. The nebular spectra of these events looked like a normal SN~Ia \citep{Tucker19,nebularspec18oh} and had no evidence for H/He , [\ion{Ca}{2}], or [\ion{O}{1}] emission. 
With the \textit{James Webb Space Telescope}, we can further explore the diversity in the mid/far-infrared spectra of SNe~Ia at late phases \citep[e.g., see][]{Kwok2023, DerKacy2023}, and search for further clues for the source of the difference between SNe~Ia with and without early bumps.

There have now been a handful of normal SNe~Ia with detected early excess flux \citep{dimitriadis2018k2, 2017cbv, Ashall2022, 2021aefxHosseinzadeh}.  Notably, two of these SNe (SNe~2018oh and 2023bee) were observed by either \kepler\ or \tess.  These SNe at 49 and 32~Mpc, respectively, are also among the closest SNe~Ia observed by either observatory. The only comparably close SNe~Ia in this combined sample \citep[e.g.,][and other individual SNe discussed above]{olling2015, fausnaugh2019early, Fausnaugh2023} are SNe~2018fhw \citep{vallely2019} and 2018hib \citep{fausnaugh2019early}, at 74~Mpc and 66~Mpc, respectively.  SN~2018hib did not have any indication of an early flux excess \citep{fausnaugh2019early}.  While SN~2018fhw did not display early excess flux \citep{vallely2019}, it has an abnormal nearly linear rise and its late-time spectra had hydrogen emission from circumstellar material, indicative of companion interaction \citep{Kollmeier2019, vallely2019}.  Therefore, half of the nearby SNe~Ia observed by either \kepler\ or \tess\ have a flux excess with an additional object having other properties that indicate there may have been a flux excess if viewed from a different angle.  Considering viewing angle effects, the physical mechanism that creates early flux excesses may be ubiquitous for all SNe~Ia.

On the other hand, the detected rate of early excess drops dramatically when it comes to SNe Ia at higher redshift. \cite{Deckers2022} did a systematic analysis on the SNe~Ia sample with early coverage from ZTF, and finds that 3 out of 30 SNe~Ia with $z<0.07$ have detectable early excess, though there is large difference in data conditions such as signal-to-noise ratio and cadence between the brightest SNe~Ia sample and the ZTF sample. \cite{Fausnaugh2023} did a systematic search for the early excess features in the early light curves of 74 SNe~Ia in \tess\ Sector 1-50, only found 3 tentative candidates, and none of them are robust detection with the BIC test. The brightest SNe~Ia with early excesses also show a wide variety of excess flux morphology and brightness relative to the SN brightness, e.g., a 50\% difference in excess flux brightness between SN~2018oh and SN~2023bee, as well as differences in their early colors between SN~2017cbv, SN~2021aefx and SN~2023bee. The sample of high-cadence SN~Ia from the space and the ground has grown significantly in the last few years, but we can only take advantage by pushing toward fainter SN~Ia with better efficiency and contamination analyses that take the diversity of the bumps as well as the artifacts in the observational data fully into account.


\section*{Acknowledgement}

The authors would like to acknowledge Y.\ Murakami and M.\ Fausnaugh for useful discussions. The author would like to acknowledge the help of S.\ Lai and W.\ J.\ Hon and C.\ Onken on obtaining data with SSO-2.3m telescope.

This paper includes data collected by the TESS mission. Funding for the \tess\ mission is provided by the NASA's Science Mission Directorate. The \tess\ data presented in this paper were obtained from the Mikulski Archive for Space Telescopes (MAST) at the Space Telescope Science Institute (STScI). The specific observations analyzed can be accessed via \dataset[https://doi.org/10.17909/0cp4-2j79]{https://doi.org/10.17909/0cp4-2j79}. STScI is operated by the Association of Universities for Research in Astronomy, Inc., under NASA contract NAS5–26555. Support to MAST for these data is provided by the NASA Office of Space Science via grant NAG5–7584 and by other grants and contracts.

This work was partially supported by TESS grant 80NSSC21K0242 and NASA ADAP grant 80NSSC22K0494.
Q.W.\ is supported in part by NASA grants 80NSSC19K0112 and STScI DDRF fund.
C.D.K.\ acknowledges partial support from a CIERA postdoctoral fellowship.
C.R.A.\ and L.I.\ were supported by a grant from VILLUM FONDEN (project number 16599)
M.R.S.\ is supported by the STScI Postdoctoral Fellowship.
C.G.\ is supported by a VILLUM FONDEN Young Investigator Grant (project number 25501).
G.D.\ is supported by the H2020 European Research Council grant no. 758638.
C.A.\ acknowledges support by NASAgrants JWST-GO-02114.032-A and JWST-GO-02122.032-A.
M.R.M.\ acknowledges a Warwick Astrophysics prize post-doctoral fellowship made possible thanks to a generous philanthropic donation.
K.A. would also like to acknowledge Ian Price and Chris Lidman with their help with observations taken with the ANU 2.3-metre telescope.  The automation of the ANU 2.3-metre telescope was made possible through funding provided by the Centre of Gravitational Astrophysics at the Australian National University.

The UCSC team is supported in part by NASA grant NNG17PX03C, NSF grant AST--1815935, the Gordon \& Betty Moore Foundation, the Heising-Simons Foundation, and by a fellowship from the David and Lucile Packard Foundation to R.J.F.

The Young Supernova Experiment (YSE) and its research infrastructure is supported by the European Research Council under the European Union's Horizon 2020 research and innovation programme (ERC Grant Agreement 101002652, PI K.\ Mandel), the Heising-Simons Foundation (2018-0913, PI R.\ Foley; 2018-0911, PI R.\ Margutti), NASA (NNG17PX03C, PI R.\ Foley), NSF (AST-1720756, AST-1815935, PI R.\ Foley; AST-1909796, AST-1944985, PI R.\ Margutti), the David \& Lucille Packard Foundation (PI R.\ Foley), VILLUM FONDEN (project 16599, PI J.\ Hjorth), and the Center for AstroPhysical Surveys (CAPS) at the National Center for Supercomputing Applications (NCSA) and the University of Illinois Urbana-Champaign.

Pan-STARRS is a project of the Institute for Astronomy of the University of Hawaii, and is supported by the NASA SSO Near Earth Observation Program under grants 80NSSC18K0971, NNX14AM74G, NNX12AR65G, NNX13AQ47G, NNX08AR22G, 80NSSC21K1572 and by the State of Hawaii.  The Pan-STARRS1 Surveys (PS1) and the PS1 public science archive have been made possible through contributions by the Institute for Astronomy, the University of Hawaii, the Pan-STARRS Project Office, the Max-Planck Society and its participating institutes, the Max Planck Institute for Astronomy, Heidelberg and the Max Planck Institute for Extraterrestrial Physics, Garching, The Johns Hopkins University, Durham University, the University of Edinburgh, the Queen's University Belfast, the Harvard-Smithsonian Center for Astrophysics, the Las Cumbres Observatory Global Telescope Network Incorporated, the National Central University of Taiwan, STScI, NASA under grant NNX08AR22G issued through the Planetary Science Division of the NASA Science Mission Directorate, NSF grant AST-1238877, the University of Maryland, Eotvos Lorand University (ELTE), the Los Alamos National Laboratory, and the Gordon and Betty Moore Foundation.

This work has made use of data from the Asteroid Terrestrial-impact Last Alert System (ATLAS) project. The Asteroid Terrestrial-impact Last Alert System (ATLAS) project is primarily funded to search for near-Earth objects (NEOs) through NASA grants NN12AR55G, 80NSSC18K0284, and 80NSSC18K1575; byproducts of the NEO search include images and catalogs from the survey area. This work was partially funded by Kepler/K2 grant J1944/80NSSC19K0112 and HST GO-15889, and STFC grants ST/T000198/1 and ST/S006109/1. The ATLAS science products have been made possible through the contributions of the University of Hawaii's Institute for Astronomy, the Queen’s University Belfast, the Space Telescope Science Institute, the South African Astronomical Observatory, and The Millennium Institute of Astrophysics (MAS), Chile.

This work includes data obtained with the Swope Telescope at Las Campanas Observatory, Chile, as part of the Swope Time Domain Key Project (PI: Piro, Co-Is: Drout, Phillips, Holoien, French, Cowperthwaite, Burns, Madore, Foley, Kilpatrick, Rojas-Bravo, Dimitriadis, Hsiao). We thank A.\ Campillay and Y.\ Kong Riveros for performing the Swope observations.

This project used data obtained with the Dark Energy Camera (DECam), which was constructed by the Dark Energy Survey (DES) collaboration. Funding for the DES Projects has been provided by the U.S. Department of Energy, the U.S. NSF, the Ministry of Science and Education of Spain, the Science and Technology Facilities Council of the United Kingdom, the Higher Education Funding Council for England, the National Center for Supercomputing Applications at the University of Illinois at Urbana-Champaign, the Kavli Institute of Cosmological Physics at the University of Chicago, Center for Cosmology and Astro-Particle Physics at the Ohio State University, the Mitchell Institute for Fundamental Physics and Astronomy at Texas A\&M University, Financiadora de Estudos e Projetos, Fundacao Carlos Chagas Filho de Amparo, Financiadora de Estudos e Projetos, Fundacao Carlos Chagas Filho de Amparo a Pesquisa do Estado do Rio de Janeiro, Conselho Nacional de Desenvolvimento Cientifico e Tecnologico and the Ministerio da Ciencia, Tecnologia e Inovacao, the Deutsche Forschungsgemeinschaft and the Collaborating Institutions in the Dark Energy Survey. The Collaborating Institutions are Argonne National Laboratory, the University of California at Santa Cruz, the University of Cambridge, Centro de Investigaciones Energeticas, Medioambientales y Tecnologicas-Madrid, the University of Chicago, University College London, the DES-Brazil Consortium, the University of Edinburgh, the Eidgenossische Technische Hochschule (ETH) Zurich, Fermi National Accelerator Laboratory, the University of Illinois at Urbana-Champaign, the Institut de Ciencies de l’Espai (IEEC/CSIC), the Institut de Fisica d’Altes Energies, Lawrence Berkeley National Laboratory, the Ludwig Maximilians Universitat Munchen and the associated Excellence Cluster Universe, the University of Michigan, NSF’s NOIRLab, the University of Nottingham, the Ohio State University, the University of Pennsylvania, the University of Portsmouth, SLAC National Accelerator Laboratory, Stanford University, the University of Sussex, and Texas A\&M University.

This publication has made use of data collected at Lulin Observatory, partly supported by MoST grant 108-2112-M-008-001.

A major upgrade of the Kast spectrograph on the Shane 3~m telescope at Lick Observatory was made possible through generous gifts from the Heising-Simons Foundation as well as William and Marina Kast. Research at Lick Observatory is partially supported by a generous gift from Google.

The data presented here were obtained (in part) with ALFOSC, which is provided by the Instituto de Astrofisica de Andalucia (IAA) under a joint agreement with the University of Copenhagen and NOT.

Based in part on observations obtained at the Southern Astrophysical Research (SOAR) telescope, which is a joint project of the Minist\'{e}rio da Ci\^{e}ncia, Tecnologia e Inova\c{c}\~{o}es (MCTI/LNA) do Brasil, the US NSF's NOIRLab, the University of North Carolina at Chapel Hill (UNC), and Michigan State University (MSU).

YSE-PZ was developed by the UC Santa Cruz Transients Team with support from NASA grants NNG17PX03C, 80NSSC19K1386, and 80NSSC20K0953; NSF grants AST-1518052, AST-1815935, and AST-1911206; the Gordon \& Betty Moore Foundation; the Heising-Simons Foundation; a fellowship from the David and Lucile Packard Foundation to R.J.F.; Gordon and Betty Moore Foundation postdoctoral fellowships and a NASA Einstein fellowship, as administered through the NASA Hubble Fellowship program and grant HST-HF2-51462.001, to D.O.J.; and a National Science Foundation Graduate Research Fellowship, administered through grant No.\ DGE-1339067, to D.A.C.

Parts of this research were supported by the Australian Research Council Centre of Excellence for All Sky Astrophysics in 3 Dimensions (ASTRO 3D), through project number CE170100013.

We thank the Precision Observations of Infant Supernova Explosions (POISE) collaboration for contributing late-time photometry of 2023bee to this paper.

\section*{Software and third party data repository citations} \label{sec:cite}

%

\facilities{\tess, DECam, ATLAS, PS1,  LCO, Swope, Swift, Thacher, Lulin, NOT, Shane, YAO:2.4m, FTS, ATT, APO:3.5m}


\software{astropy \citep{2013A&A...558A..33A,2018AJ....156..123A},
\texttt{TESSreduce} \citep{tessreduce}, 
Matplotlib \citep{matplotlib}, 
SciPy \citep[][]{2020SciPy-NMeth}, 
NumPy \citep{2020NumPy-Array}, 
pyphot \citep[][]{fouesneaumorgan_2022}, 
lmfit \citep{lmfit},
YSE-PZ \citep{Jones2021,Coulter2022, Coulter2023},
SNCosmo \citep{sncosmo}}



\begin{figure}[t!]
    \centering
    \hspace*{-0.2in}
    \includegraphics[width=0.7\textwidth]{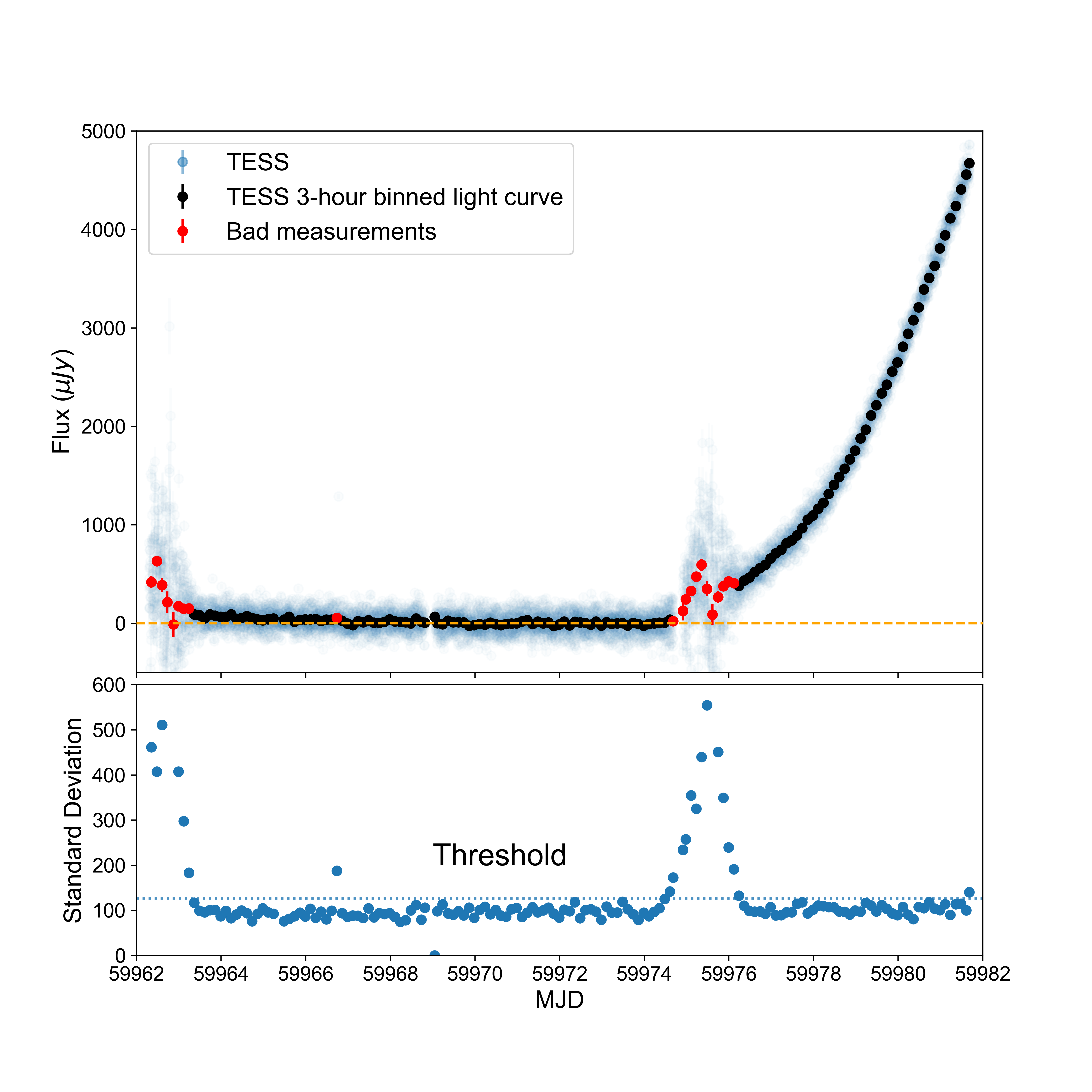}
    \caption{\textbf{Top:} the raw (blue) and 3-hour binned (black) \tess\ light curve of SN~2023bee. The red points denote the bad measurements, determined by a threshold characterized by their high uncertainty. Notice that before MJD 59969 there is a subtle trend in background flux, but between MJD 59969 and 59974 the background flux flattens out. \textbf{Bottom:} the standard deviations of light curve bins. The horizontal dashed line denotes the threshold defined as $\Bar{\sigma} + 5\sigma_\sigma$ for data between MJD 59969.2 and 59974 when the background flux is stable. 
    }
    \label{fig:cleanlc}
\end{figure}

\begin{figure}[t!]
    \centering
    \hspace*{-0.2in}
    \includegraphics[width=0.9\textwidth]{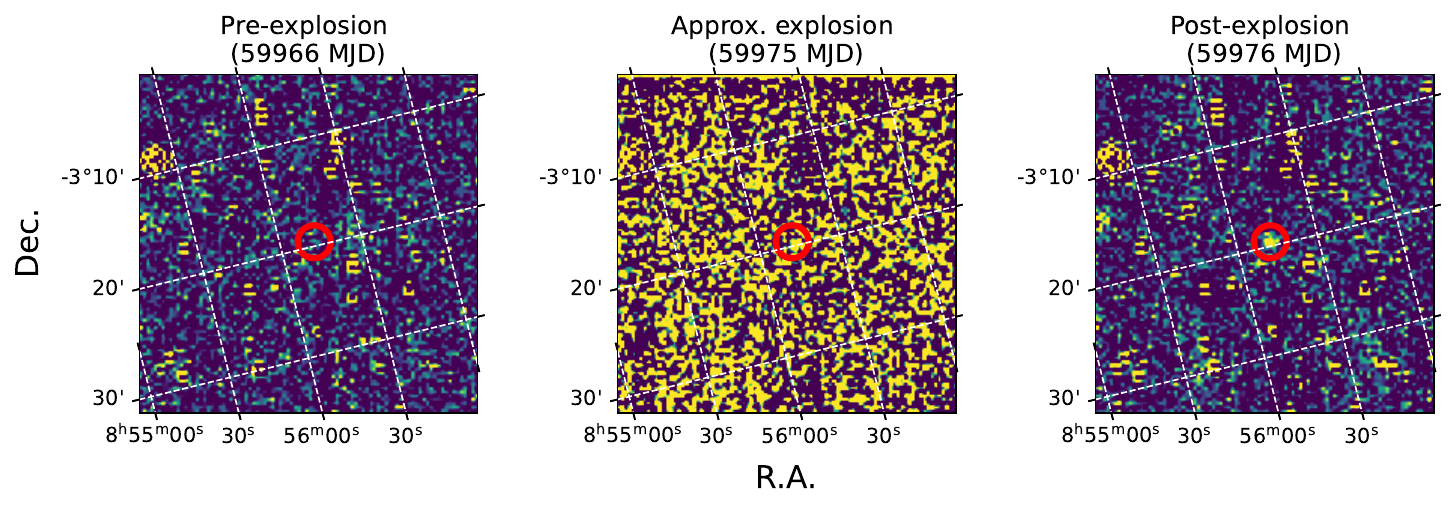}
    \caption{\tess\ image during the pre-explosion phase (left), around the time of first light when the background is saturated (middle), and during the rising phase when the background become stable again (right). Plots are in the same color scale and yellow are the saturated points.}
    \label{fig:tessimage}
\end{figure}

\appendix
\label{appendix}
Fig.~\ref{fig:cleanlc} shows the raw and 3-hour binned \tess\ light curve, and clearly the data around MJD 59975 become severely noisy. Fig.~\ref{fig:tessimage} shows the images before, during, and after this period, revealing that there is severe saturation in this part of the chip during this time.
As can be seen from the bottom panel of Fig~\ref{fig:cleanlc}, those bad datapoints tend to have large scatter and thus can be characterised by the large sample standard deviation of flux measurements within each bin. Thus, we apply a cut based on the standard deviation of the binned data to remove those bad data in \tess\ light curve. Firstly, we selected a time range between MJD 59969.2 and 59974 in which the raw \tess\ light curve is stable to estimate the baseline flux and associated standard deviation $\sigma_i$ for each 3-hour $i$th bin. 
We calculate the mean $\Bar{\sigma}$ and standard deviation $\sigma_\sigma$ of the $\{\sigma_i\}$ in this time range, and then we set the threshold for bad data as $\sigma_i > \Bar{\sigma} + 5\sigma_\sigma$. The threshold is plotted as the horizontal dotted line in the bottom panel of Fig~\ref{fig:cleanlc}, and visually it serves as a good cut in between the noisy and normal datapoints.

\begin{table*}
\centering
 \begin{tabular}{c c c c c} 
 \hline
 Obs Date & MJD & \tablenotemark{a}Phase & Telescope & Instrument \\ 
 (UT) &  & (Rest-Frame Days) & &\\ [0.3ex] 
 \hline\hline
  2023-02-01 & 59976.76& -15.74 & Lijiang-2.4m & YFOSC\\
2023-02-02	& 59977.56 & -14.94 & FTS & FLOYDS-S\\
2023-02-05	& 59980.28 & -12.23 & APO-3.5m & KOSMOS\\
2023-02-08	& 59984.04 & -8.50 & NOT & ALFOSC\\
2023-02-10	& 59985.26 & -7.29 & SOAR & GOODMAN\\
2023-02-11	& 59986.96 & -5.59 & NOT & ALFOSC\\
2023-02-18	& 59993.59 & 1.01  & SSO-2.3m & WiFeS\\
2023-02-19	& 59994.58 & 1.99  & SSO-2.3m & WiFeS\\
2023-02-20	& 59995.47 & 2.88 & SSO-2.3m & WiFeS\\
2023-02-20	& 59995.96 & 3.36 & NOT & ALFOSC\\
2023-02-27	&  60002.01 & 9.38 & NOT & ALFOSC\\
2023-02-27	&  60002.99 & 10.36 & NOT & ALFOSC\\
2023-03-08	& 60011.43 & 18.76 & SSO-2.3m & WiFeS\\
2023-03-09	& 60012.05 & 19.38 & NOT & ALFOSC\\
2023-03-10	& 60013.47 & 20.79 & SSO-2.3m & WiFeS\\
2023-03-19	& 60022.47 & 29.74 & SSO-2.3m & WiFeS\\
2023-03-25	& 60028.42 & 35.66 & SSO-2.3m & WiFeS\\
2023-04-12	&  60046.21 & 53.36 & Shane & KAST\\
2023-04-23	&  60057.20 & 64.29 & Shane & KAST\\
2023-04-30	&  60064.18 & 71.24 & Shane & KAST\\
\hline
\end{tabular}
\tablenotetext{a}{Phases relative to B-band maximum on MJD 59992.58 according to the SALT3 fit.}
\caption{Log of Spectroscopic Observations of SN~2023bee. }
\label{tab:spec}
\end{table*}

\bibliography{reference}{}
\bibliographystyle{aasjournal}



\end{document}